\documentclass[twoside,twocolumn,9pt]{article}
\usepackage{extsizes}
\usepackage[super,sort&compress,comma]{natbib} 
\usepackage[version=3]{mhchem}
\usepackage[left=1.5cm, right=1.5cm, top=1.785cm, bottom=2.0cm]{geometry}
\usepackage{balance}
\usepackage{mathptmx}
\usepackage{sectsty}
\usepackage{graphicx} 
\usepackage{lastpage}
\usepackage[format=plain,justification=justified,singlelinecheck=false,font={stretch=1.125,small,sf},labelfont=bf,labelsep=space]{caption}
\usepackage{float}
\usepackage{fancyhdr}
\usepackage{fnpos}
\usepackage{nccmath, mathtools}
\usepackage[english]{babel}
\addto{\captionsenglish}{%"References"
}
\usepackage{array}
\usepackage{droidsans}
\usepackage{charter}
\usepackage[T1]{fontenc}
\usepackage[usenames,dvipsnames,table,xcdraw]{xcolor}
\usepackage{setspace}
\usepackage[compact]{titlesec}
\usepackage[hidelinks]{hyperref}
\usepackage[make-links=true]{acro}
\usepackage{enumerate}
\usepackage{multirow}
\definecolor{cream}{RGB}{222,217,201}
\usepackage{breakurl}

\usepackage{titling}
\title{Accumulate}

\acsetup{make-links=true, pages/display = all}
\DeclareAcronym{AML}{short=AML , long=Anti-money laundering}
\DeclareAcronym{ADI}{short=ADI , long=Accumulate digital identifier}
\DeclareAcronym{BME}{short=BME , long=Burn and mint equilibrium}
\DeclareAcronym{BPT}{short=BPT , long=Binary Patricia Trie}
\DeclareAcronym{BVN}{short=BVN , long=Block validator network}
\DeclareAcronym{BVNN}{short=BVNN , long=Block validator network node}
\DeclareAcronym{DAO}{short=DAO , long=Decentralized autonomous organization}
\DeclareAcronym{DAG}{short=DAG , long=Directed acyclic graph}
\DeclareAcronym{DDII}{short=DDII , long=Decentralized digital identity and identifier}
\DeclareAcronym{DN}{short=DN , long=Directory network}
\DeclareAcronym{DPoS}{short=DPoS , long=Delegated proof of stake}
\DeclareAcronym{DSN}{short=DSN , long=Data server network}
\DeclareAcronym{IoT}{short=IoT , long=Internet of things}
\DeclareAcronym{IIoT}{short=IIoT , long=Industrial Internet of things}
\DeclareAcronym{KYC}{short=KYC , long=Know your customer}
\DeclareAcronym{NFT}{short=NFT , long= Non-fungible token}
\DeclareAcronym{PoS}{short=PoS , long=Proof of stake}
\DeclareAcronym{PoW}{short=PoW , long=Proof of work}
\DeclareAcronym{SDAO}{short=SDAO , long=Sponsored decentralized autonomous organization}
\DeclareAcronym{SMT}{short=SMT , long=Stateful Merkle Tree}
\DeclareAcronym{TPS}{short=TPS , long=Transactions per second}
\DeclareAcronym{Tx}{short=Tx , long=Transaction}
\DeclareAcronym{s}{short=URL , long=Uniform resource locator}

\begin{document}

\pagestyle{fancy}
\thispagestyle{plain}
\fancypagestyle{plain}{
%%%HEADER%%%
\renewcommand{\headrulewidth}{0pt}
}
%%%END OF HEADER%%%

%%%PAGE SETUP - Please do not change any commands within this section%%%
\makeFNbottom
\makeatletter
\renewcommand\LARGE{\@setfontsize\LARGE{15pt}{17}}
\renewcommand\Large{\@setfontsize\Large{12pt}{14}}
\renewcommand\large{\@setfontsize\large{10pt}{12}}
\renewcommand\footnotesize{\@setfontsize\footnotesize{7pt}{10}}
\makeatother

\renewcommand{\thefootnote}{\fnsymbol{footnote}}
\renewcommand\footnoterule{\vspace*{1pt}% 
\color{cream}\hrule width 3.5in height 0.4pt \color{black}\vspace*{5pt}} 
\setcounter{secnumdepth}{5}

\makeatletter 
\renewcommand\@biblabel[1]{#1}            
\renewcommand\@makefntext[1]% 
{\noindent\makebox[0pt][r]{\@thefnmark\,}#1}
\makeatother 
\renewcommand{\figurename}{\small{Fig.}~}
\sectionfont{\sffamily\Large}
\subsectionfont{\normalsize}
\subsubsectionfont{\bf}
\setstretch{1.125} %In particular, please do not alter this line.
\setlength{\skip\footins}{0.8cm}
\setlength{\footnotesep}{0.25cm}
\setlength{\jot}{10pt}
\titlespacing*{\section}{0pt}{4pt}{4pt}
\titlespacing*{\subsection}{0pt}{15pt}{1pt}
%%%END OF PAGE SETUP%%%

%%%FOOTER%%%
\fancyfoot{}
\fancyfoot[LO,RE]{\vspace{-7.1pt}\includegraphics[height=9pt]{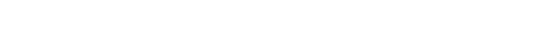}}
\fancyfoot[CO]{\vspace{-7.1pt}\hspace{12.8cm}{\small Accumulate Whitepaper, v1.0, 04/2022}}
\fancyfoot[CE]{\vspace{-7.8pt}\hspace{-12.7cm}{\small Accumulate Whitepaper, v1.0, 04/2022}}
\fancyfoot[RO]{\footnotesize{\sffamily{~\textbar  \hspace{2pt}\thepage}}}
\fancyfoot[LE]{\footnotesize{\sffamily{\thepage~\textbar\hspace{3.45cm}}}}
\fancyhead{}
\renewcommand{\headrulewidth}{0pt} 
\renewcommand{\footrulewidth}{0pt}
\setlength{\arrayrulewidth}{1pt}
\setlength{\columnsep}{6.5mm}
\setlength\bibsep{1pt}
%%%END OF FOOTER%%%

%%%FIGURE SETUP - please do not change any commands within this section%%%
\makeatletter 
\newlength{\figrulesep} 
\setlength{\figrulesep}{0.5\textfloatsep} 
\newcommand{\topfigrule}{\vspace*{-1pt}% 
\noindent{\color{cream}\rule[-\figrulesep]{\columnwidth}{1.5pt}} }
\newcommand{\botfigrule}{\vspace*{-2pt}% 
\noindent{\color{cream}\rule[\figrulesep]{\columnwidth}{1.5pt}} }
\newcommand{\dblfigrule}{\vspace*{-1pt}% 
\noindent{\color{cream}\rule[-\figrulesep]{\textwidth}{1.5pt}} }
\makeatother
%%%END OF FIGURE SETUP%%%

%%%TITLE, AUTHORS AND ABSTRACT%%%
\twocolumn[
  \begin{@twocolumnfalse}
    {\Huge \thetitle \hfill \raisebox{0pt}[0pt][0pt] {\includegraphics[height=55pt]{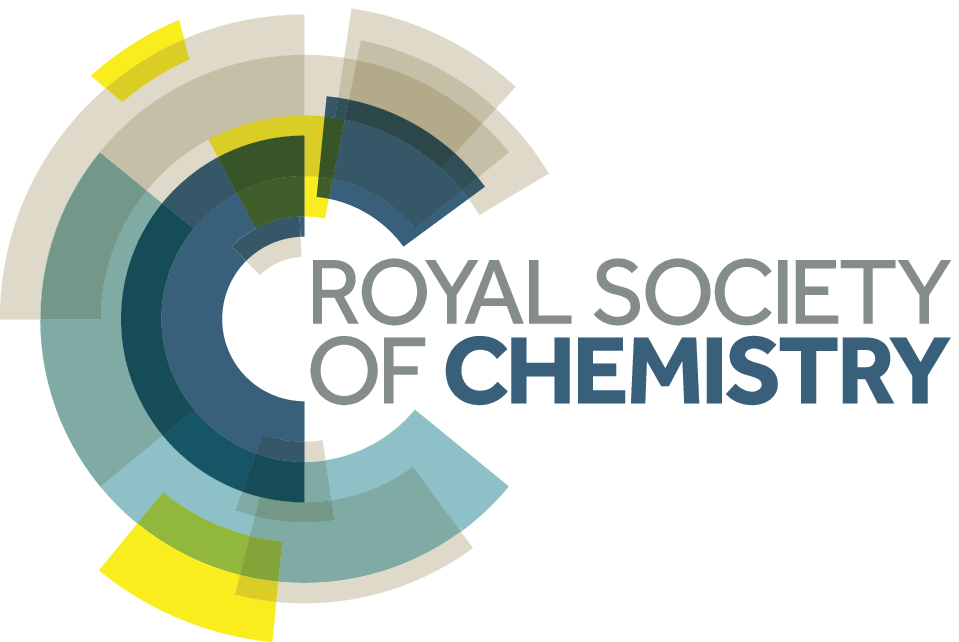}}\\[1ex]
    \includegraphics[width=18.5cm]{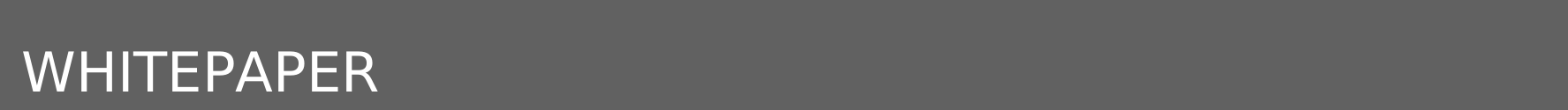}}\par
    \vspace{2.55em}
    \sffamily
    \hspace{-0.1cm}\begin{tabular}{m{4.4cm} p{13.6cm} }
    
      \hspace{-0.3cm}\begin{tabular}{l}\vspace{1.5em}\\
            \footnotesize Version: 1.0  \\
            \footnotesize Date: 12 Apr 2022 \\
            \footnotesize DOI: 10.48550/arXiv.2204.06878 \\
            %\footnotesize DOI: 10.6084/m9.figshare.19586515 \\
        \end{tabular}

        & \noindent\LARGE{\textbf{Accumulate: An identity-based blockchain protocol with cross-chain support, human-readable addresses, and key management capabilities}} \\ %Article title
        \vspace{0.3cm} & \vspace{0.3cm} \\
        
        \hspace{-0.3cm}\begin{tabular}{l}\footnotesize Publisher: Inveniam DeFi Devs, Inc. \\ 
        \footnotesize Contact: paulsnow@defidevs.io\\ \end{tabular} & \noindent\large{Kyle Michelson, Anjali Sridharan, Umut Can Çabuk, Ethan Reesor, Ben Stolman, Drew Mailen, Dennis Bunfield, Jay Smith, and Paul Snow.} \\\\%Author names

        \hspace{-0.3cm}\begin{tabular}{l}\footnotesize License: CC-BY-NC-ND\ \\ \end{tabular} & \noindent\normalsize{The Accumulate Protocol (“Accumulate”) is an identity-based, Delegated Proof of Stake (\acs{DPoS}) blockchain designed to power the digital economy through interoperability with Layer-1 blockchains, integration with enterprise tech stacks, and interfacing with the World Wide Web. Accumulate bypasses the trilemma of security, scalability, and decentralization by implementing a chain-of-chains architecture in which digital identities with the ability to manage keys, tokens, data, and other identities are treated as their own independent blockchains. This architecture allows these identities, known as Accumulate Digital Identifiers (\acs{ADI}s), to be processed and validated in parallel over the Accumulate network. Each \acs{ADI} also possesses a hierarchical set of keys with different priority levels that allow users to manage their security over time and create complex signature authorization schemes that expand the utility of multi-signature transactions. A two token system provides predictable costs for enterprise users, while anchoring all transactions to Layer-1 blockchains provides enterprise-grade security to everyone.} \\%The abstract
    
    \end{tabular}
 \end{@twocolumnfalse}
\vspace{1cm}
]
%%%END OF TITLE, AUTHORS, AND ABSTRACT%%%

%%%FONT SETUP
\renewcommand*\rmdefault{bch}\normalfont\upshape
\rmfamily
\section*{}
\vspace{-1cm}

%%%FOOTNOTES%%%
%\footnotetext{\textit{$^{a}$~Address, Address, Town, Country. Fax: XX XXXX XXXX; Tel: XX XXXX XXXX; E-mail: xxxx@aaa.bbb.ccc}}
%\footnotetext{\textit{$^{b}$~Address, Address, Town, Country. }}

%\footnotetext{\dag~Electronic Supplementary Information (ESI) available: [details of any supplementary information available should be included here]. See DOI: 00.0000/00000000.}

%\footnotetext{\ddag~Additional footnotes to the title and authors can be included \textit{e.g.}\ `Present address:’ or `These authors contributed equally to this work’ as above using the symbols: \ddag, \textsection, and \P. Please place the appropriate symbol next to the author’s name and include a \texttt{\textbackslash footnotetext} entry in the correct place in the list.}%%%END OF FOOTNOTES%%%

%%%%%%%%%%%%%%%%%%%%%%%%%%%%%%%%%%%%%%%%%%%%%%%%%%%%%%%%%%%%%%%%%%%%%
%%%%%%%%%%%%%%%%%%%%%%%% MAIN TEXT %%%%%%%%%%%%%%%%%%%%%%%%%%%%%%%%%%
%%%%%%%%%%%%%%%%%%%%%%%%%%%%%%%%%%%%%%%%%%%%%%%%%%%%%%%%%%%%%%%%%%%%%

\setcounter{tocdepth}{2}  
{\small \balance \tableofcontents } \newpage

\section{Introduction}

%Some intro to introduction

\subsection{Motivation \& Problem Statement}

Designing data storage and management methods that provide a high level of security and efficiency is a challenge of high importance to blockchain protocols. An ideal platform includes desirable characteristics such as fast read/write speeds (low block and transaction times), low transaction costs, scalability, security, pathways for migrations of legacy systems, and easy management of data access and authorizations. These standards have proved difficult to meet for existing blockchain protocols, with computationally-expensive smart contracts often lacking the efficiencies and security required for scaling. Accumulate addresses these requirements through its unique identity-based architecture, with the following key innovations that set it apart from other protocols:
%REFERENCE needed for claim that smart contracts have scaling issues
\begin{itemize}
    \item Use of Decentralized Digital Identity and Identifiers (\acs{DDII}) as the atomic unit of the blockchain.
    \item Chain-of-chains architecture for network partitioning and linear scalability.
    \item Key Books for easy management of credentials and access.
    \item Synthetic transactions for inter-chain communication. 
    \item Scratch Accounts for efficiently managing consensus.
\end{itemize}
 
This document begins by providing a high-level overview of each of these key innovations, followed by a technical description of the overall system architecture. These key innovations will be revisited with a more technical emphasis on how each fits into the overall architecture of the protocol, and how they enable a variety of real-world use cases.

\subsection{Inheritance from Factom}
The Factom protocol was launched in 2014 as an alternative to the direct use of the Bitcoin blockchain for the management and organization of data, making it one of the earliest Layer-2 blockchains \cite{Snow2018}. At the time, blockchain technology was being explored as a means to reliably and securely store data without requiring trust between unfamiliar parties or the involvement of a centralized third party to validate transactions and maintain a record of the transaction history. However, Bitcoin’s verification process required an entire ledger to validate individual pieces of data, a computationally expensive method. Factom’s founding developers recognized that Bitcoin was inconvenient and unsuitable for enterprise data solutions with its simple linear blockchain format and limited capacity.

Factom was designed to utilize the security and reliability of the Bitcoin protocol while making notable improvements to its efficiency in record storage applications. As a Layer-2 protocol, Factom was able to commit to the Bitcoin blockchain in near real-time, secured by the consensus mechanism of the Factom protocol itself \cite{Snow2018}. Factom was designed with a unique chain-of-chains architecture in which a set of chains were linked together into a directory layer. Data for applications built on the protocol was contained in specific data chains, allowing for faster indexing and reduced computational and storage costs for organizations.

The Factom protocol took security one step further by anchoring the results of this consensus to the Bitcoin blockchain every 10 minutes using a data-efficient Merkle tree. Merkle trees \cite{Merkle1979} were used to organize large amounts of data entered into Factom chains to be secured onto the Bitcoin blockchain with a single data point. This provided users with an immutable history of their transactions and allowed them to track and sync with only a small portion of the Factom Protocol to run their application. Factom also introduced several innovations for more efficient scaling. For example, sharding was simplified due to the separation of token transactions and computation from the data layer. An efficiently sharded data blockchain allows for continuous and real-time securing of data at the level demanded by enterprise customers \cite{Snow2018}. 

With its highly competitive speeds and transaction times, Factom was able to forge several successful partnerships in the public and private sectors. Many of these were groundbreaking implementations of blockchain technology, such as a pilot program with the Department of Energy (DOE) to secure the energy grid of Texas, USA; a grant from the Department of Homeland Security (DHS) protect the authenticity of data collected by Internet of Things (\acs{IoT}) devices, and an award by the Gates Foundation to store and organize medical records in developing countries \cite{Scottnov2016}. Determined to push the boundaries of the protocol’s innovation, Factom’s developers began re-envisioning its utility and creating what would eventually become the Accumulate Protocol. With this new design, the founders of Accumulate were committed to preserving Factom’s unique and innovative aspects while improving on its ease of use, mass appeal, and scalability.
%REFERENCES TO NEWS ARTICLES FOR DOE, DHS, GATES

\subsection{Key Innovations}
Since Accumulate is a continuation of the Factom protocol, a comparison between Accumulate and Factom is useful. Accumulate introduces many key innovations while carrying over Factom's core features, making the protocol significantly more robust and versatile.

\begin{itemize}
\item \textbf{The Identity Paradigm:} 
While most protocols today are tied to the paradigm of public-private keys as a means of storing and accessing records,
Accumulate is the first blockchain organized around decentralized, self-sovereign digital identifiers. Creating user-friendly, application-friendly, maintainable identifiers is an important challenge addressed by this protocol. The core organizational structure used in Accumulate is the Accumulate digital identity/identifier/domain (\acs{ADI}) \cite{litepaper}. The use of \acs{ADI}s has allowed Accumulate to bypass the ‘Scalability Trilemma’, a commonly-observed trade-off between decentralization, security, and scalability \cite{Monte2020}. %Examples of this include Bitcoin prioritizing security over scalability with longer block delays, EOS choosing scalability at the cost of decentralization, etc.
%REFERENCE%

Accumulate overcomes the trilemma by organizing itself around \acs{ADI}s,  where each \acs{ADI} defines its own state that is independent of other \acs{ADI}s. \acs{ADI}s are distributed over a set of Tendermint networks, each of which is referred to as a Block Validator Network (\acs{BVN}). Each \acs{BVN} is responsible for a particular set of \acs{ADI}s, and as the Accumulate Network grows, more \acs{BVN}s can be added to scale and maintain high throughput.   
\item \textbf{Parallel Processing:}
 Each \acs{ADI} is treated as having its own state and set of accounts and
chains. Each \acs{ADI} is updated independently from the state of other \acs{ADI}s. This highly branched design allows more parallel processing and greater freedom to distribute the responsibilities of processing state 
across many networks. This allows for fast transaction finalization and short block times.

\item \textbf{Key Management:} 
Any organization that handles data has a
system of access control for the various levels of authorized
users on their system. Accumulate reproduces this capability
within a blockchain protocol by providing each \acs{ADI} with the
ability to create Key Books for defining these access controls.
Key Books define how decisions are to be made and even how
the Key Book itself is managed. The Key Book holds Key Pages
that define the keys and authorization schemes to be used in decision making. The Keys and the authorization schemes specified in a Key Page can be managed and 
modified, eliminating any need to reissue accounts just because an organization must shift responsibilities/privileges over time. This flexibility in key management gives users the tools they
need to dynamically model, manage, and maintain the 
dynamic nature of blockchain applications.

\item \textbf{Synthetic Transactions:}
Since each \acs{ADI} has its own state, transactions that are routed to an \acs{ADI} must be processed independently of all other \acs{ADI}s in the network. This becomes challenging when two or more \acs{ADI}s are involved in a transaction. Accumulate's solution is to generate an additional transaction that performs settlement within an \acs{ADI}. Transactions generated by the protocol in response to transactions initiated by a user are referred to as Synthetic Transactions. 

In practice, a Synthetic Transaction is sent to the \acs{BVN} that manages the destination \acs{ADI} after the source \acs{ADI} initiates a transaction. Debits to the source \acs{ADI} are handled directly by the user's transaction, while credits to the destination account are handled by the Synthetic Transaction. A Synthetic Transaction is sent every time an \acs{ADI} interacts with another \acs{ADI}, which provides each \acs{BVN} with a balance of every \acs{ADI} that has debited or credited an \acs{ADI} within its own network. This way, an \acs{ADI} does not need to query any \acs{BVN} in the Accumulate network to verify the balance of a source \acs{ADI}.

% REPLACED:
%Accumulate mitigates this risk through the use of latency between the validation and recording of transactions, as well as before the outcome of a transaction changes the state of the blockchain through the use of Synthetic Transactions. Accumulate directly processes transactions submitted by users; however, these cannot change the state of the blockchain. Instead, the user transactions create synthetic transactions that are processed by the protocol. Any transaction (user or synthetic) is restricted to modifying at most one identity; thus, transactions that involve different identities may be processed in parallel.

\item \textbf{Scratch Accounts:}
Another key vulnerability in many blockchains is the off-blockchain processes that are needed to arrive at an entry on the blockchain. Developers are faced with a trade-off between the expense and difficulty of showing their work on the blockchain and the lack of accountability when the work of coming to consensus is done off-chain. Accumulate provides scratch accounts, which reduce the cost of using the blockchain for consensus building. However, the data availability of scratch accounts is limited. Scratch accounts allow processes to provide cryptographic proof of validation and process without overburdening the blockchain. 
\end{itemize}

\subsection{Integrations}
Blockchain protocols each tend to exhibit their own niche design and function. Accumulate is first and foremost a system of identity management capable of modeling complex organizational structures and relationships. This makes it ideal for integration with various other protocols, improving the accessibility of their blockchain technologies in a more diverse variety of use cases. 

The Accumulate protocol does not natively write smart contracts but it does support various smart contract roll-ups. This allows Accumulate to track the state and validity of contracts stored on third-party chains. An organization is then able to process smart contracts written across a variety of Layer-1 protocols such as Ethereum, Solana, and Tezos. In the near future, when vendors may rely on a preferred Layer-1 protocol to conduct their operations (such as using smart contacts for supply-chain tracking), Accumulate will have the capability to manage this multitude of blockchains.   

The Accumulate ecosystem also has room for integration with Layer-0 protocols, which aim to connect the hundreds of existing unique protocols. Known as ‘blockchains of blockchains’, protocols such as Cosmos and Polkadot\cite{Wood2016} have made great improvements towards interoperability, or the ability to make transactions across two different protocols efficiently. Accumulate can then be utilized to manage the transferred asset under the identity or \acs{ADI} of the buyer and to continue to track the movement of the asset across multiple identities. This allows for the custodial transfer of assets through multiple organizations, using whatever protocols they consider ‘native’.

The use of Accumulate to create an ecosystem has several clear benefits. One of Accumulate’s most valuable contributions is its security model that allows for rapid response to security breaches. By moving away from the paradigm of a 1:1 relationship between key and asset, Accumulate decouples the concept of identity and security. This ensures that even if a transaction’s private keys are compromised, an organization can set its own authority of trusted individuals to prevent an incorrect transfer. This is especially important in the case of assets that are repeatedly transferred across various protocols when there is an increased risk that private keys could become exposed.

\section{System Overview}
At the systems level, Accumulate is optimized for parallel processing, linear scaling, and efficiency of state. Parallel processing is achieved by partitioning the network into multiple validator networks that process transactions for only a fraction of accounts. Additional validator networks can be added to linearly scale the network as usage increases. Efficiency of state is realized through Accumulate’s chain-of-chains architecture that organizes transactions into hierarchies of summary hashes and allows a user to validate transactions on a mobile device. This chapter explains the architectural basis of these features and provides a high-level overview of the Accumulate protocol. A detailed description of the Accumulate architecture is provided in Chapter \ref{architecture_overview}.

\subsection{Introduction}
In a traditional blockchain, transactions are hashed using cryptographic algorithms like SHA-256\cite{Appel2015} and organized into a data structure called a Merkle tree\cite{Merkle1979}. Pairs of hashes are concatenated until a single hash remains (the Merkle root). The Merkle tree is stored within a block and connected to other blocks in the network by including the hash of the previous block in the current block header. This data structure creates a block ‘chain’ that maintains the history of the network. Multiple full nodes, each containing a complete copy of the blockchain, participate in block validation and maintain the state of the network. If a node is compromised and a transaction within a block is modified, then the Merkle roots of all subsequent blocks will change and the network will reject the fraudulent block.

The Accumulate protocol has a fundamentally different architecture that is organized around accounts rather than blocks. Each account is treated as an independent chain and managed as a continuously growing Merkle tree while blocks are treated as synchronization points for all chains in the network. The indexing of these points allows Accumulate to provide 1 second blocks for finalization and 12 hour blocks for synchronization of the historical ledger. This is because blocks are defined by indexing the Merkle Trees of chains rather than building independent Merkle Trees rooted to blocks. Shorter block times are constrained by the speed at which the Tendermint consensus layer can validate transactions. Longer block times require less state and allow mobile devices to work as Lite Nodes that can validate a subset of transactions relevant to the account owner. 

\subsection{Validation}
\begin{figure}[!h]
 \centering
 \includegraphics[scale=0.35]{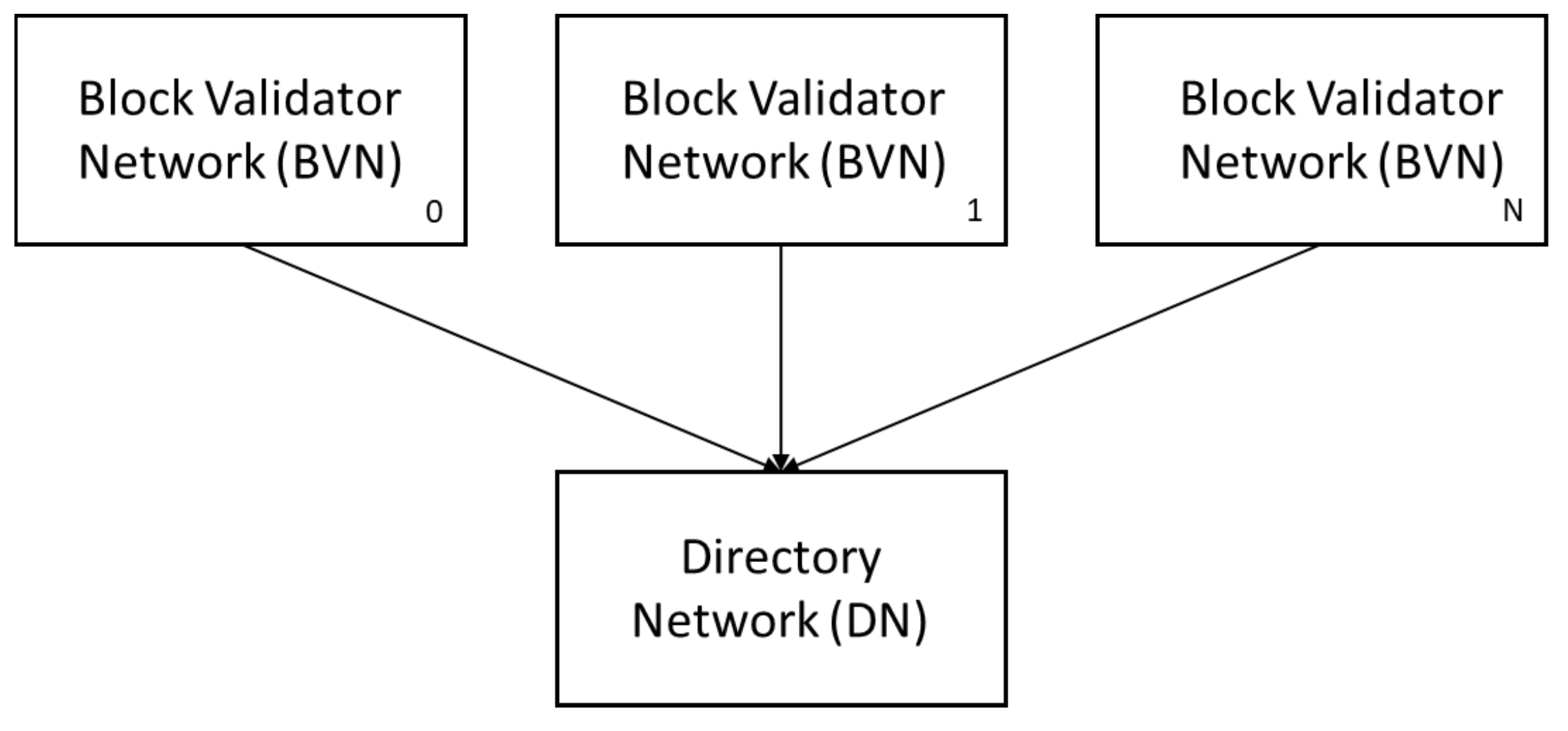}
 \caption{System overview.}
 \label{fig:overview}
\end{figure}

Networks of Tendermint nodes, called Block Validator Networks (\acs{BVN}s), validate transactions for an account. Each account is assigned to a particular \acs{BVN}, and each \acs{BVN} can process transactions for thousands of accounts. Adding more \acs{BVN}s allows the Accumulate network to scale linearly as utilization grows. As shown in Figure \ref{fig:overview}, root hashes from each \acs{BVN} are fed into a larger network of Tendermint nodes called the Directory Network (\acs{DN}). The \acs{DN} produces a final root hash that can be ‘anchored’ into another blockchain. \acs{DN} root hashes will be regularly anchored into Layer-1 blockchains like Bitcoin and Ethereum by inserting the \acs{DN} root hash as data contained within a transaction. Anchoring essentially buys the security of a larger network at a cost that is independent of the number of transactions. For example, a root hash derived from 10,000 transactions can be anchored into Bitcoin at the cost of a single Bitcoin transaction.

\begin{figure*}[tb]
 \centering
 \includegraphics[scale=0.35]{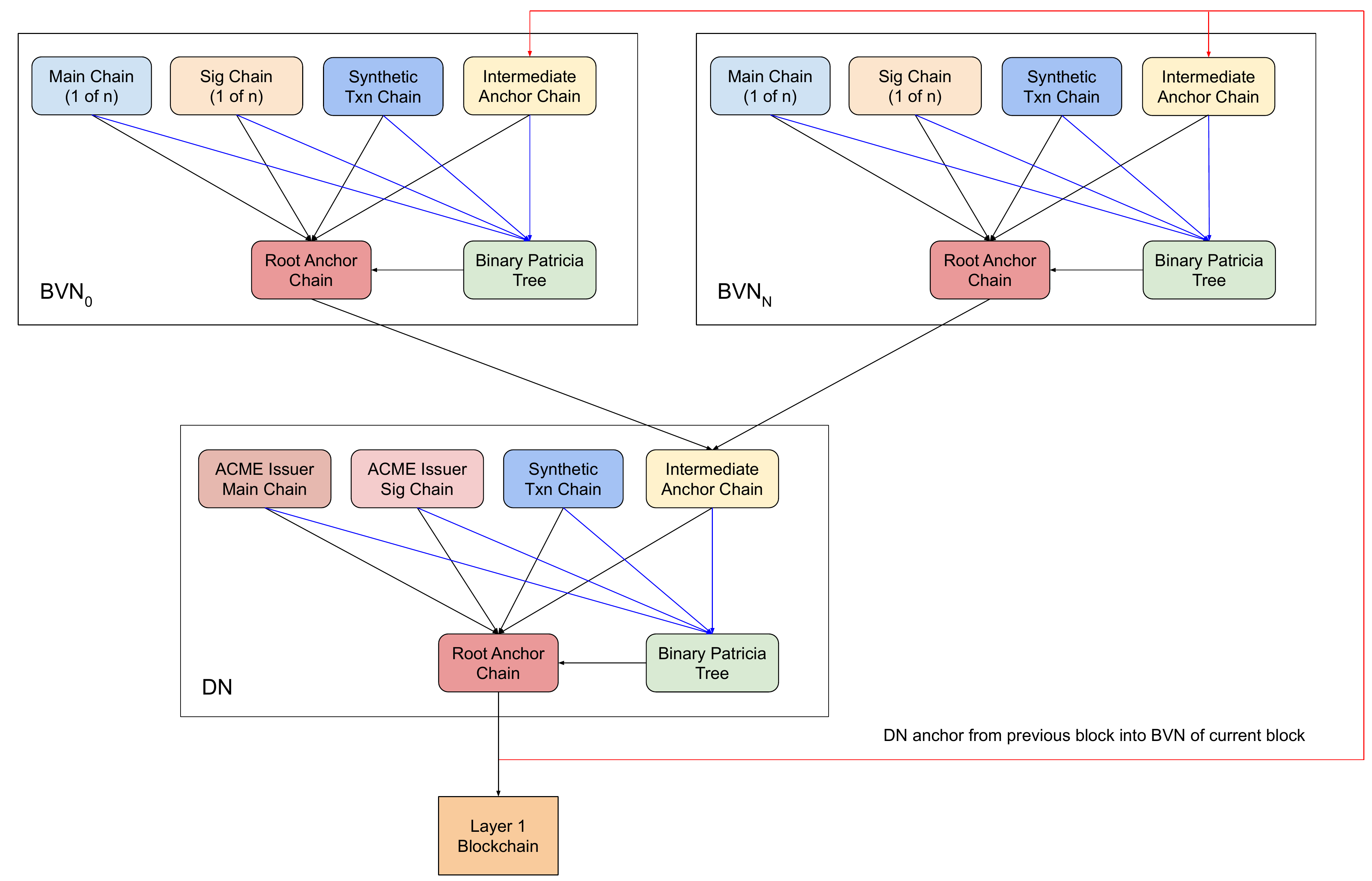}
 \caption{Chains-of-chains architecture of Accumulate.}
 \label{fig:chains}
\end{figure*}

\subsection{Chain of Chains Architecture}
Inside the \acs{DN} and each \acs{BVN} is an interconnected network of chains responsible for collecting signatures, synchronizing Tendermint nodes within a \acs{BVN}, communicating between \acs{BVN}s, collecting roots, and anchoring roots to other blockchains. These chains and their functions are summarized below and illustrated in Figure \ref{fig:chains}, which is an expanded view of Figure \ref{fig:overview}.

\begin{itemize}
    \item Signature (Sig) Chain: Collects signatures for a period of approximately 2 weeks, after which transactions that have not met the $m-of-n$ signature threshold are discarded.
    \item Main Chain: Records transactions in both the origin account and any account modified by the transaction once a transaction has met the signature threshold(s) and has been executed.
    \item Synthetic Transaction Chain: Synthetic transactions are produced when a transaction needs to update accounts across multiple identities. The Synthetic Transaction Chain provides cryptographic proof that a synthetic transaction was actually produced by a particular \acs{BVN}.
    \item Binary Patricia Trie (\acs{BPT}): Collects hashes representing the current state and history of every account in the \acs{BVN} or \acs{DN}.
    \item Root Anchor Chain: Collects an anchor once per block from every account chain and system chain updated during the block. For every major and minor block, the Root Anchor Chain produces a root hash for the \acs{BVN} or \acs{DN}. The root hash of the \acs{DN} is the root hash of the entire Accumulate network.
    %\item \acs{BVN} Root Anchor Chain: Collects anchors from the Synthetic Transaction Chain of a \acs{BVN} and the Transaction and Signature Chains of each account within a \acs{BVN} to produce a root hash every major and minor block.
    %\item \acs{DN} Root Anchor Chain: Collects anchors from the Synthetic Transaction Chain of the \acs{DN} and the Signature and main chains of the ACME token issuer to produce a root hash for every major and minor block.
    \item Intermediate Anchor Chain: Within the \acs{DN}, Intermediate Anchor Chains collect an anchor from the Root Anchor Chain of every \acs{BVN}, once per block. Within a \acs{BVN}, it collects an anchor from the \acs{DN} Root Anchor Chain once per block.
\end{itemize}

%Each account has a Signature Chain for collecting signatures, and a Main Chain for recording signed and validated transactions. These are collectively referred to as account chains because they interact directly with the account holder. As illustrated in Figure 2, each \acs{BVN} can contain multiple account chains. However, the Signature Chain and Main Chain of an account must reside on a single \acs{BVN}.
Each account has a Signature Chain for collecting signatures, and a Main Chain for recording signed and validated transactions. These are collectively referred to as account chains, as each one is tied to a specific account. As illustrated in Figure \ref{fig:chains}, each \acs{BVN} can contain multiple accounts, each of which has a Signature Chain and a Main Chain. Each \acs{BVN} (and the \acs{DN}) has a Synthetic Transaction Chain, a Root Anchor Chain, and one or more Intermediate Anchor Chains. These are collectively referred to as system chains, as they are not tied to any specific account and are used by the system as a whole.

%ER: This does not mention blocks because it was getting confusing. Add a separate explanation of blocks, or find a way to explain “all of this happens once per block” in this paragraph.
Within a \acs{BVN} or the \acs{DN}, every chain is anchored into the Root Anchor Chain. Therefore, every chain can be considered a side chain of the Root Anchor Chain. The Root Anchor Chain of each \acs{BVN} is also anchored into the \acs{DN}'s Intermediate Anchor Chain. Since the \acs{DN}'s Intermediate Anchor Chain is anchored into the \acs{DN}'s Root Anchor Chain, every \acs{BVN} and thus every account in the network can be considered a side chain of the \acs{DN}.

%KM: Maybe expand upon in a new section
%Partitioning the Accumulate network into multiple \acs{BVN}s allows for parallel processing of transactions, but as more \acs{BVN}s are added, coordinating consensus between \acs{BVN}s can become computationally intensive. To resolve this issue, each \acs{BVN} has a Synthetic Transaction chain that efficiently communicates transactions between accounts such that each \acs{BVN} has a complete record of anything that is debited or credited to it. 

%The Intermediate Anchor Chain of each \acs{BVN} is responsible for coordinating consensus with the \acs{DN} and forming a sequential blockchain. This is achieved by anchoring the \acs{DN} root of the previous block into the Intermediate Anchor Chain of each \acs{BVN}. The Intermediate Anchor Chain of the \acs{DN}, meanwhile, is responsible for collecting roots for every \acs{BVN} in the network.

Each of these chains is organized as a continuously growing Merkle tree from which a root hash, or anchor, can be produced every block. This creates a chain-of-chains architecture where every chain within a \acs{BVN} or the \acs{DN} can be considered a side chain of its respective \acs{BVN}/\acs{DN} Root Anchor Chain. A key/value pair representing the state of an account is constructed for each account and added into the Binary Patricia Trie (\acs{BPT}). Thus, the \acs{BPT} is responsible for recording the current state of the Accumulate network. This architecture and its applications, such as Scratch Accounts, Managed Transactions, and Delegated Transactions, are discussed in greater detail in Chapter \ref{architecture_overview}.

\subsection{Account Architecture}
Accumulate accounts are designed to improve the user experience, efficiently organize data, and integrate with web apps, mobile devices, and enterprise tech stacks. Accumulate supports the following types of accounts:

\begin{itemize}
    \item Lite Token Account: Traditional blockchain address whose URL contains a public key hash and human-readable suffix denoting the token or data type held by the account.
    \item Lite Data Account: Provides compatibility with Factom. Public data account that can be used for collaboration.
    \item Accumulate Digital Identifier (\acs{ADI}): The primary unit of organization within Accumulate. Besides Lite Accounts, all other accounts must belong to an \acs{ADI}.
    \item Key Book: Secures accounts and provides advanced key management. Belongs to an \acs{ADI}. See Chapter \ref{key_management}.
    \item Key Page: Organizes keys within a Key Book.
    \item \acs{ADI} Token Account: Holds and exchanges tokens. Belongs to an \acs{ADI}.
    \item \acs{ADI} Data Account: Records data entries. Belongs to an \acs{ADI}.
    \item Scratch Account: An \acs{ADI} Data or Token Account with limited data availability. Transactions older than approximately two weeks will not be available.
    %\item Lite Accounts: Traditional blockchain addresses whose URL contains a public key hash and human-readable suffix denoting the token or data type held by the account. There are Lite Data and Lite Token accounts.
    %\item Accumulate Digital Identifiers (\acs{ADI}s): Completely human-readable URLs with a hierarchical structure and key management capability. There are \acs{ADI} data, token, staking, and scratch accounts that can operate independently or in combination with other \acs{ADI}s under the management of one or more \acs{ADI}s.
    %\item Key Pages: Key management tools that define the set of Keys required to validate a transaction. Key pages are contained within a Key Book and contain a number of keys specified by a user. However, Key Books and Keys are not considered accounts because they lack main and Signature chains.
\end{itemize}

Each account is uniquely identified by a URL. When tokens are sent to a Lite Token Account URL, the account is created if it does not already exist. A new user will get started with Accumulate by creating a key and having a trusted party send tokens to the corresponding Lite Token Account URL. For example, the user may pay USD or BTC to an exchange, which will send tokens from the exchange’s Lite Token Account or \acs{ADI} Token Account to the URL specified by the user. Lite Data Accounts are created through a similar mechanism, but they are only capable of writing data. Since Lite Data accounts do not specify keys, anyone with the address is able to write data to the account. This feature may be useful to an organization but is of little value to a potential hacker. A more detailed description of Lite Accounts is provided in Chapter \ref{lite_accounts}.

% Alternatively, the user could ask a friend (trusted party?) to use their account to send tokens to the address.

%Lite Token Accounts are free to create on the Accumulate network, but they only exist once a transaction using the native ACME token is sent to the protocol from an external account. An \acs{ADI} or ACME token holder with tokens held on an exchange or inside an external wallet could send ACME to the Accumulate network and specify a Lite Token Account creation through the mobile app or command-line interface. In future releases, Lite Token Accounts will be able to hold other types of tokens besides ACME. Lite Data Accounts are created similarly, but they are only able to write data. Since Lite Data accounts do not specify keys, anyone with the address is able to write data to the account. This feature may be useful to an organization but is of little value to a potential hacker. A more detailed description of Lite Accounts and their structure is found in Chapter \ref{lite_accounts}.

\acs{ADI}s are versatile accounts that perform many of the same functions as Lite Accounts in addition to being able to issue their own tokens, stake native ACME tokens, participate in complex authorization schemes, and manage other accounts. Most of these operations are enabled by hierarchical sets of keys, including Key Books and Key Pages, which are introduced in the next section and discussed in detail in Chapter \ref {key_management}. \acs{ADI}s are purchased using Credits, a non-transferable utility token with a fixed USD value that is created when ACME tokens are burned. Credits are used to pay for all operations on the Accumulate network such as token issuance, account creation, and updates to keys. An introduction to Accumulate's dual token model is provided in Chapter \ref{tokenomics}.

%ER: Move to Key Management
%Each account is linked at creation to a Main Key Book. Signatures of transactions originating from an account are authorized against the account’s Main Key Book. Any number of Key Books can belong to an \acs{ADI}, and each account in the \acs{ADI} can be linked to any Key Book in the \acs{ADI}.

%Transactions must be signed/authorized by $m$ of $n$ keys from a Key Page.
%Key Pages contain $m$ of $n$ keys that are used to sign transactions. The number of Keys is specified by a user, where $m = n = 1$ specifies a single signature transaction and $m \geq n > 1$ specifies a multi-signature transaction. An account can contain multiple Key Pages within a Key Book, and each Key Page has its own Signature chain and main chain. 

\subsection{Key Architecture}

%ER: Should this be subsumed into Key Management? Or can it be shortened to a summary of Key Management? It seems kind of redundant at the moment.

%Accumulate has a hierarchical key management structure where Key Books contain Key Pages, and Key Pages specify m-of-n keys required to approve a transaction. An \acs{ADI} contains at least one Account Key Book and one optional Manager Key Book. A Manager Key Book has the authority to approve or reject a transaction submitted by the \acs{ADI}. If a transaction is rejected, a manager can suggest a modified transaction on the \acs{ADI}’s Signature chain. An Account Key Book is maintained by the user and is generally responsible for creating transactions.

Accumulate has a hierarchical key management structure where Key Books contain Key Pages and Key Pages contain keys that are authorized to sign transactions. Each Key Page specifies $m$ of $n$ where $m$ is the number of unique, authorized signatures required to approve transactions (i.e., the signature threshold) and $n$ is the total number of keys on the page.

Accounts are linked at creation to a Main Key Book and an optional Manager Key Book. A signature is authorized only if the signing key corresponds to a key in one of the pages of either the Main Key Book or the Manager Key Book (if specified). When creating a transaction, the user selects a Key Page from the Main Key Book of the origin of the transaction to use for the transaction. Signatures of transactions originating from an account are authorized against the account’s Main Key Book. Any number of Key Books can belong to an \acs{ADI}, and each account in the \acs{ADI} can be linked to any Key Book in the \acs{ADI}.

The number of Key Pages within a Key Book is specified by the user. Each Key Page is assigned a different priority level where a Key Page with a higher priority can make changes to any Key Page with a lower priority (e.g., change $m-of-n$ requirements). A user can also specify any number of Key Books from external \acs{ADI}s as the owners of any number of Key Pages within their Key Book. External Key Books can participate in transactions within the Key Book of another \acs{ADI} and even make changes to the security of Key Pages with lower priority. These are called Delegated Transactions because they are made on behalf of another party. Delegated Transactions and Managed Transactions are formally introduced in Chapters \ref{delegated} and \ref{managed}, respectively. Key management is explained in more detail in Chapter \ref{key_management}.

\section{Identity Management}
Participation in the Accumulate network occurs through Accumulate Digital Identifiers (\acs{ADI}s) and Lite Accounts, similar to how participation in other blockchains occurs through a wallet or address. \acs{ADI}s give users access to the full range of features provided by the Accumulate network, including smart contracts, off-chain consensus building, and dynamic key management. Lite Accounts are a ‘lite’ version of \acs{ADI}s that, despite their comparatively limited utility and flexibility, may appeal to users who simply want to send and receive tokens and maintain a record of their transactions.

\subsection{Lite Accounts} \label{lite_accounts}
The Accumulate protocol supports two types of Lite Accounts for basic management of tokens and data. Lite Token Accounts are similar to a traditional blockchain address and are used for sending and receiving tokens. The token type and authorization scheme of a Lite Token Account is specified by the user, but it cannot be changed after the account is created. However, a user can create multiple accounts in their wallet to manage different tokens.

Lite Data Accounts were designed to be compatible with Factom and help its users migrate to Accumulate. For this reason, Lite Data Accounts are constructed a bit differently than other accounts and mirror the way that Factom creates hashes and account IDs. Anyone who knows the URL of a Lite Data Account can append data similar to how anyone who knows the URL of a Lite Token Account can deposit tokens. However, data entries cannot be overwritten. Spamming a Lite Data Account is possible, but this type of attack is disincentivized because every data transaction incurs a small fee. Since Lite Data Accounts have limited utility outside of migration from Factom to Accumulate, they will not be covered in the following sections.

\subsubsection{Generating a Lite Token Account}
Anyone can generate a Lite Account by arranging for tokens to be sent to a Lite Token Account address corresponding to a public key that they own. \acs{ADI}s, in contrast, can only be created by a sponsor through the spending of Credits issued through the Accumulate protocol. A user without an \acs{ADI} to use as a sponsor can buy ACME tokens on an exchange, create a Lite Account for ACME tokens, purchase credits on the network, and sponsor the creation of their own \acs{ADI}. They can then use their \acs{ADI} to sponsor the creation of additional \acs{ADI}s for themselves or others.

A Lite Token Account consists of an identity and token URL. The Accumulate wallet automatically derives the identity by concatenating the SHA-256 hash and checksum of a user’s public key. The token URL (e.g., bob/token) is specified by the user and appended to the identity. A Lite Token Account is created by the first transaction sent to its URL. This transaction may originate from an exchange or an external wallet (e.g., Metamask). When Accumulate  receives the transaction, it will automatically create a Lite Account if the Lite Account URL is valid and a Lite Account does not already exist for that URL.

To spend tokens in the Lite Token Account, a user will initiate a transaction, hash and sign it with their private key, attach their public key, and submit this to the Accumulate network. The network will hash the public key and compute a checksum, verify that the concatenated hash and checksum are identical to the Lite Token Account provided by the user, then decrypt the signature with their public key and verify the withdrawal.

\subsubsection{Lite Token Account URLs}
The URL format of a Lite Account is generalized as: \\

\noindent acc://<keyHash><checkSum>/<tokenUrl>. \\

The <keyHash> is the first 20 bytes of the SHA-256 sum of the account’s public key, encoded as hexadecimal. The <checkSum> is the last 4 bytes of the SHA-256 sum of the lower case <keyHash>, also in hexadecimal. The following example illustrates the process of creating a Lite ACME Token Account from a public/private key pair:\\

{\noindent $PrivK$: \indent b270eaaa57e5d4d808a9766f64b340aa655481c298288\\\indent\indent\indent\indent238e5b1b3561fb80b27

\noindent $PubK$: \indent 023e6165e349c2822089ab042b3a885ca54a0907e237e\\\indent\indent\indent\indent8bfb5bd2aa96885966f35 \\

\noindent Compute the SHA-256 hash of the public key:\\

\noindent $H(PubK)$: 818D7C1F69E7BEBCE54FE087F44D86D14279100D2\\\indent\indent\indent\indent\indent{EEA690AC3847AE1B9A14237} \\

\noindent Trim this hash to the first 20 bytes (odds of a random match is 1 in $10^{48}$) and convert to lowercase: \\

\noindent $H'(PubK)$: 818d7c1f69e7bebce54fe087f44d86d14279100d \\

\noindent Compute checksum (last 4 bytes of the SHA-256 hash of trimmed hash) and convert to lowercase: \\

\noindent$C(H'(PubK)):$ 904a336d \\

\noindent Concatenate (20 bytes of) the hash, the checksum, a slash, and the token URL: \\

\noindent $URL$:\indent acc://818d7c1f69e7bebce54fe087f44d86d14279100d90\\\indent\indent\indent4a336d/acme} \\

The length of a Lite Token Account is fixed at 48 characters, where the hash of the public key occupies 40 characters and the appended checksum adds an additional 8 characters. Thus, the length of a Lite Token Account URL is fixed at 49 characters (including 1 character for the slash) plus the length of the token URL. When submitting a transaction to a Lite Token Account, a user can choose whether to append the acc:// prefix. If a prefix is not added by the user, it will be automatically added by the network.

Appending a checksum to the hash of the public key prevents the irreversible loss of tokens if the public key hash is copied incorrectly or derived from another algorithm. When the server receives a transaction, it converts the Lite Token Account identity into bytes, takes the first 20 bytes representing the key hash, recalculates the checksum from the first 40 hexadecimal characters, and verifies that they match. If the checksum is incorrect, the transaction fails and the tokens are returned to the sender. This process functions as a built-in verification mechanism, which is particularly important for Lite Token Accounts since they are not comprised of human-readable text and are therefore easier to mistype.

%\subsubsection{Security}
%The names of \acs{ADI}s and their accounts are chosen by the user. If a user creates an \acs{ADI} based solely on the key hash from their Lite Account, this has the potential to confuse the user since some accounts will be Lite Accounts while others will be tied to \acs{ADI}s. %The creation of a token registry for all tokens associated with a public key hash will block its reuse and also provide a user with a record of their token accounts. While these accounts cannot be managed like \acs{ADI}s, a registry will help users keep track of their assets and minimize the chance that tokens in a particular account will be forgotten. A token registry creates a record of the public key hash in the Patricia Trie \cite{Jung2002} as well as a record of individual token accounts in the Merkle Tree that can be queried to provide a list of tokens held by an account at any point in time.

%\subsubsection{Lite Token Accounts }
%(also talk about $m = n = 0$ so anyone can write to the account. Like if they want to hash data to your \acs{ADI} token account. Probably similar for Lite Data Accounts but ask Ethan

%\subsubsection{Lite Data Accounts}

\subsection{ADIs}\label{adis}
%\subsubsection{Introduction}
 Accumulate Digital Identifiers (\acs{ADI}s) are URL-based digital identities with hierarchical key sets that can manage data, tokens, and other identities. They can also be used to govern more complicated operations such as token issuance, off-chain consensus building, and complex authorization schemes (e.g. multi-signature). While key hierarchies are useful for organizing priorities, identity hierarchies are useful for organizing objects. Objects might include different token types held by a user, different departments within an organization, or different data types collected by an array of \acs{IoT} sensors. How a company organizes its departments, financial records, or data structures can be reproduced in the organization of \acs{ADI}s, sub-ADIs, their data and token accounts, and their directories as defined below: 

\begin{itemize}
    \item \acs{ADI}: A digital identifier that governs data, tokens, and identities. An \acs{ADI} manages these assets using a set of hierarchical keys that it owns or shares with other \acs{ADI}s.
    \item Sub-\acs{ADI}: An nth generation \acs{ADI} that is created within another \acs{ADI}.
    \item \acs{ADI} Data/Token Account: A terminal Data or Token Account belonging to an \acs{ADI}.
    \item \acs{ADI} Scratch Account: An \acs{ADI} Token/Data Account with a transaction history that is only retained for 2 weeks.
    \item \acs{ADI} Token Issuer: Defines a type of token that can be issued to \acs{ADI} Token Accounts and Lite Token Accounts.
    \item Directory: An optional cataloging structure that organizes sub-ADIs and accounts but does not manage their contents.
\end{itemize}

\subsubsection{Design}

Any number of accounts or sub-ADIs can be nested within an \acs{ADI}. Accounts and sub-ADIs are independent of each other and possess their own sets of keys with which they can manage their assets. The parent \acs{ADI} possesses an administrative key set that can add, delete, transfer, or modify the security of its accounts or sub-ADIs. Data and token accounts are terminal, meaning that they cannot have nested accounts. However, sub-ADIs and accounts can be nested within another sub-ADI.

\subsubsection{ADI Data and Token Accounts}
\acs{ADI} Data and Token Account URLs have the general format <prefix>://<ADI>/<directory>/<account> where the prefix $acc://$ specifies the Accumulate blockchain, $ADI$ specifies the top-level identity in control of the URL, $directory$ specifies a particular type of account, and $account$ specifies data or tokens. Several examples of data and token accounts are provided in the table below for the hypothetical \acs{ADI} `bank'. Note that both cryptocurrencies and tokenized assets may be considered token accounts.

\begin{table}[!htb]
\caption{Format for \acs{ADI} Data and Token accounts.} \small
\label{adi1}
\begin{tabular}{l|l|l}
\hline
\textbf{ADI}         & \textbf{acc://bank}       & \textbf{acc://bank}     \\ \hline
\textbf{Description} & Data                      & Tokens                  \\ 
\textbf{Directory}   & d                         & t                       \\ 
\textbf{Account-1}   & acc://bank/d/accounts     & acc://bank/t/loans      \\ 
\textbf{Account-2}   & acc://bank/d/investments  & acc://bank/t/securities \\ 
\textbf{Account-3}   & acc://bank/d/transactions & acc://bank/t/realestate \\ 
\textbf{Account-N}   & acc://bank/d/nth         & acc://bank/t/nth     \\ \hline
\end{tabular}
\end{table}

Directories are optional, and we anticipate that their utility will be mostly confined to enterprise applications where a greater organization of large numbers of accounts and sub-ADIs is needed. The directory label is defined by the user and can be single or multi-character. For the sake of clarity, data, tokens, and identities are given the labels $d$, $t$, and $i$. However, an investment firm with \acs{ADI} `firm' may create data accounts with directories $r$ and $c$ for residential and commercial properties, while a cryptocurrency trader and non-fungible token (\acs{NFT}) art collector may create token accounts with directories $NFT$ and $stable$ to denote NFTs and stablecoins. Any label can be used so long as the object type is specified within the wallet. However, a directory label cannot be changed after it is created.

% \begin{table}[htb] 
% \caption{Table Caption} \small
% \label{tab:adi2}
% \begin{tabular}{l|l|l}
% \hline
% \textbf{{ADI}}         & \textbf{acc://firm}     & \textbf{acc://user}    \\ \hline
% \textbf{Description} & Data                    & Data                   \\
% \textbf{Directory}   & Variable                & Variable               \\
% \textbf{Sub-ADI-1}   & acc://firm/r/properties & acc://user/nft/art     \\
% \textbf{Sub-ADI-2}   & acc://firm/c/sales      & acc://user/stable/USDT \\ \hline
% \end{tabular}
% \end{table}

\subsubsection{Organization of Sub-ADIs}

Sub-\acs{ADI}s are organized differently from data and token accounts in that any number of sub-ADIs can be nested within another sub-ADI, which can extend infinitely in both the vertical and horizontal direction. This allows a user to create an arbitrarily complex directory structure to model anything from a file directory to a corporate hierarchy. In the example below, different departments within a bank are represented by different Sub-\acs{ADI}s. The human resources department is further subdivided into sub-ADIs that specify different roles within the department such as manager and director. While not shown, any of these sub-ADIs may also have a nested data or token account.

\begin{table}[!htb] 
\caption{Sub-\acs{ADI} account format.}  
\setlength{\tabcolsep}{0.36em}
\label{adi3}
\begin{tabular}{l|l} 
\hline
\textbf{ADI}         & \textbf{acc://bank} \\ \hline
\textbf{Description} & Identity            \\
\textbf{Directory}   & i                   \\
\textbf{Sub1-ADI-1}  & acc://bank/i/auditing     \\
\textbf{Sub2-ADI-1} & acc://bank/i/auditing/i/director                \\
\textbf{Sub3-ADI-1} & acc://bank/i/auditing/i/director/i/manager     \\
\textbf{SubN-ADI-1} & acc://bank/i/auditing/i/director/i/manager/nth \\ \hline
\end{tabular}
\end{table}

\subsubsection{Combining Accounts and Sub-ADIs}
Data and token accounts may also be nested within sub-ADIs. This offers an alternative strategy for organizing large numbers of data and token accounts without needing to create multiple directories. Using the example of an investment firm that manages different types of properties, we can imagine that a team specializing in residential homes may be given a different sub-ADI than a team specializing in commercial real estate. For simplicity, we refer to these combined addresses as URLs below.
 
\begin{table}[!h]
\caption{Format for complex ADIs.}
\label{adi4}
\begin{tabular}{l|l}
\hline
\textbf{ADI}         & \textbf{acc://firm}                           \\ \hline
\textbf{Description} & Variable                                      \\
\textbf{Directory}   & Variable                                      \\
\textbf{URL-1}       & acc://firm/i/residential/d/properties         \\
\textbf{URL-2}       & acc://firm/i/commercial/i/office/d/properties \\
\textbf{URL-3}       & acc://firm/i/commercial/t/securities          \\
\textbf{Sub-ADI-N}   & acc://bank/i/nth                              \\ \hline
\end{tabular}
\end{table}
 
Since accounts are terminal, it is not possible to nest sub-ADIs within accounts. While acc://firm/i/residential/d/properties is permitted, acc://firm/d/properties/i/residential is not.

\subsubsection{Creating an \acs{ADI}}
An \acs{ADI} is created after a sponsor initiates a ‘create identity' transaction on the network. The sponsor can be an organization or a user so long as they spend the required number of tokens. The validator on the sponsor’s \acs{ADI} will validate the create identity transaction, produce a synthetic transaction (those created by the protocol), and push that out to the network to be processed in a later block. The network where the new identity resides will receive the synthetic transaction from the Block Validator Network (\acs{BVN}) and validate it against the current state of the network. 

\begin{figure}[tb]
 \centering
 \includegraphics[scale=0.36]{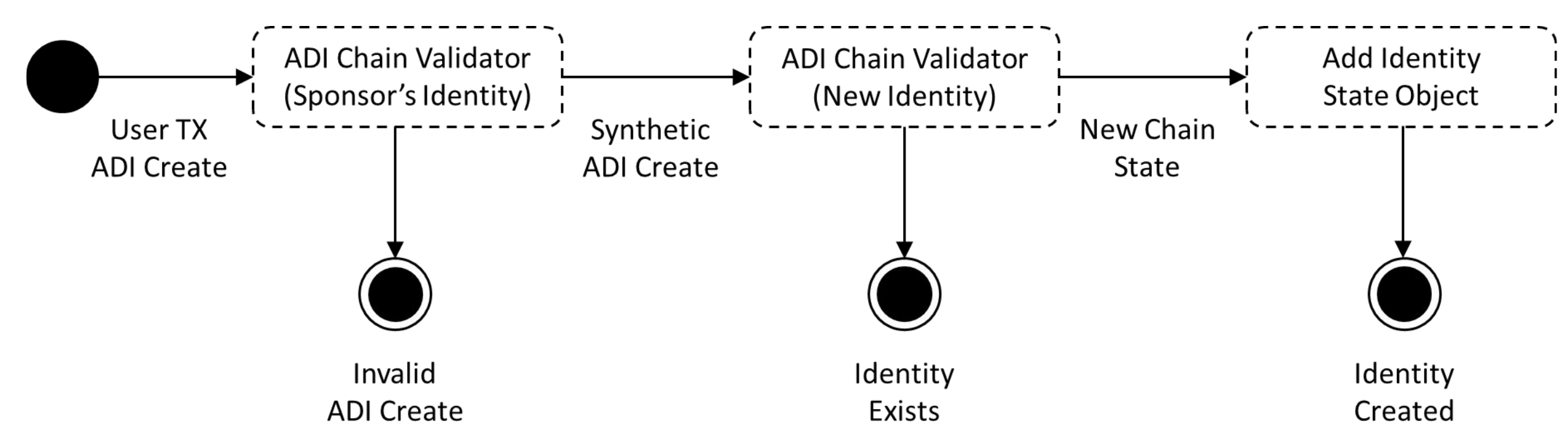}
 \caption{Creating an \acs{ADI}.}
 \label{fig:adi}
\end{figure}

A successful identity creation requires 1) that signatures for the synthetic transaction must be valid and verified against the Directory Network (\acs{DN}) receipt and 2) that the identity does not already exist on the network. If the above criteria are met, an identity state object is created, as illustrated in Figure \ref{fig:adi}. For an introduction to \acs{BVN}s and DNs, please refer to the Litepaper\cite{litepaper}. Synthetic transactions will be covered in more detail in future technical documents.
%Changed Figure 1 to Figure 3

\begin{figure*}[t]
 \centering
 \includegraphics[scale=0.65]{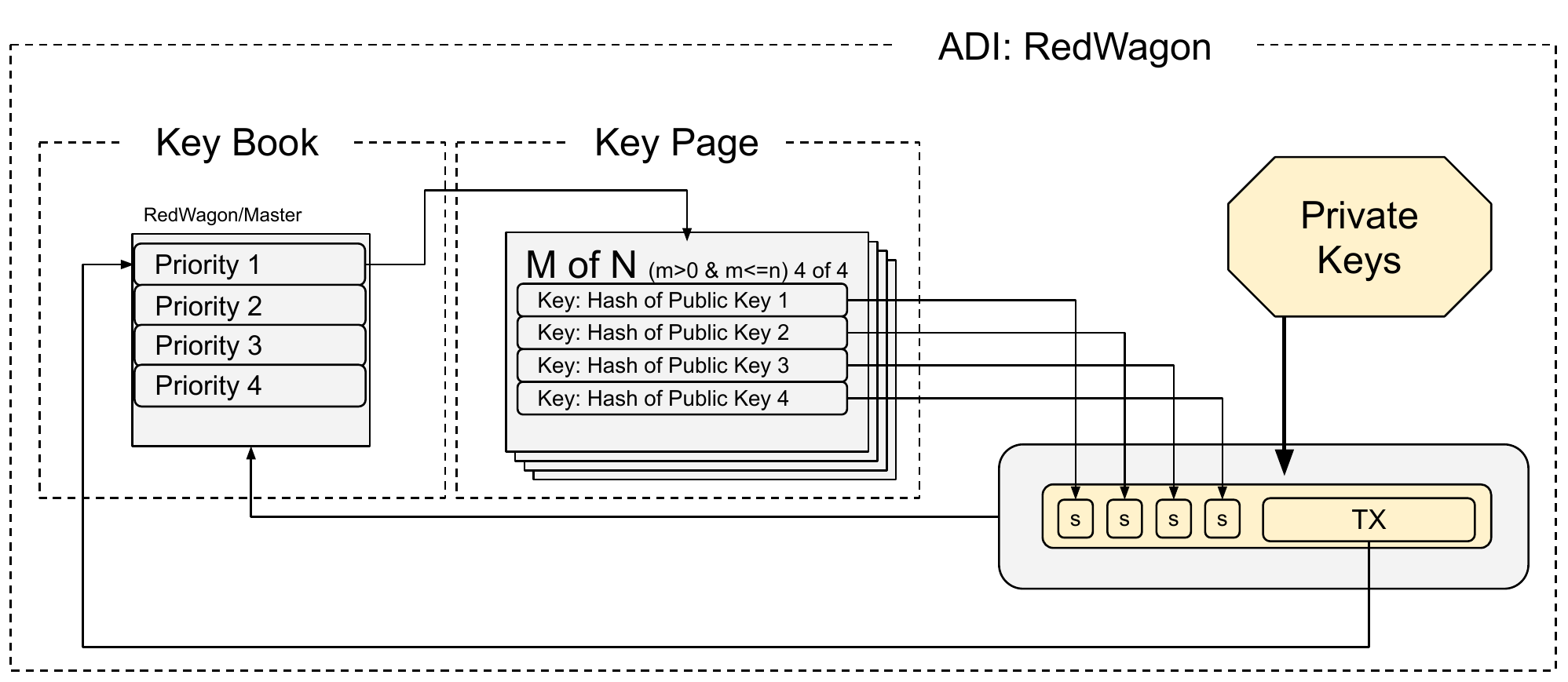}
 \caption{Key management scheme example.}
 \label{fig:key_m}
\end{figure*}
\subsubsection{Routing an \acs{ADI}}
The Accumulate network uses URLs to reference user accounts that are managed by an \acs{ADI}. The URL consists of the following characteristics:\\\\
acc://{ADI}/{Path}\\\\
Where ${ADI}$ refers to the user’s identity on the Accumulate network and $Path$ refers to accounts that are managed by an \acs{ADI}. More accurately, the \acs{ADI} is a Root Identity that is used to route transactions to \acs{BVN}s. Each account and sub-ADI within a Root Identity is handled by one \acs{BVN} that is determined after deriving the SHA-256 hash (represented by $H$ function) of the \acs{ADI} and hashing this a second time to create an address:\\\\
$AdiChainId$: H(lowerCase(\acs{ADI}))\\\\
Both the \acs{ADI} and account can be hashed such that the URL acc://RedWagon/AcmeTokens will be internally represented by the following:\\

{\noindent $H$(redwagon) and $H$(redwagon/AcmeTokens), respectively.}\\

\acs{ADI}s are evenly distributed among \acs{BVN}s, similar to how large data sets on social media platforms are evenly distributed to their servers. Network routing is determined by the first eight bytes of the SHA-256 hash of the \acs{ADI}’s account ID. The address corresponding to a particular BVN is determined by the following:\\ \\
$Address$:\indent uint64($H$($AdiChainId$) [0:7]) \\\\
where $[0:7]$ takes the first 8 bytes, and $int$ interprets those bytes as a number. The end result is a random address that can be processed internally and distributed to a particular BVN.

%\subsubsection{\acs{ADI} Token Account}
%\subsubsection{\acs{ADI} Data Account}
%\subsubsection{\acs{ADI} Scratch Account}
%\subsubsection{\acs{ADI} Staking Account}
%\subsubsection{\acs{ADI} Token Issuer Account}

\section{Key Management} \label{key_management}

Security and authorization in Accumulate are implemented with Key Books. Each \acs{ADI} must contain at least one Key Book but multiple Key Books may be specified. A Key Book specifies a set of public key hashes and Key Book Authorities that are organized and prioritized as Key Pages.

\subsection{Organization}

The Key Pages within a Key Book are organized in order of priority, where the first Key Page is the highest priority and the last Key Page is the lowest priority. However, it must be noted that the priority of a Key Page is only relevant when managing the Key Book. Each Key Page can have one or more entries, each consisting of a public key hash and/or a Key Book Authority URL.

\subsection{Management}

The available operations for managing keys are:

\begin{itemize}
    \item Creating an additional Key Book within an \acs{ADI}.
    \item Creating an additional Key Page within a Key Book.
    \item Updating the public key hash of an entry in a Key Page.
    \item Updating a Key Page.
\end{itemize}

The authorization rules for updating the public key hash of an entry in a Key Page are unique. This particular type of transaction is always single-signature and can only be used to update the entry corresponding to the signer. If the transaction is signed with a simple signature, then the only entry that can be updated is the one with a key hash that matches the signer. If the transaction is signed with a Key Book Authority, then the only entry that can be updated is the one owned by that Key Book Authority.

All other modifications of a Key Page are done via updating the Key Page where all of the normal authorization rules apply.

\subsection{Securing Accounts}

Every account must specify a Main Key Book. The Main Key Book of an account is responsible for securing the account by authorizing signatures of transactions originating from that account. The Main Key Book is set when the account is created and cannot be modified thereafter. When an \acs{ADI} is created, an initial Key Book is created along with the \acs{ADI}, which becomes the \acs{ADI}’s Main Key Book. When an account is created, any Key Book within the \acs{ADI} can be used as its Main Key Book. By default, accounts are created with the same Main Key Book as the \acs{ADI} to which they belong. While all accounts are secured by a single Key Book by default, the user has the option to create a new Key Book for every account.

Each account may also specify a Manager Key Book, which serves as a second layer of authorization. A Manager Key Book can be set at creation, but it can also be modified after the fact unlike the Main Key Book. If an account specifies a Manager Key Book in addition to its Main Key Book, then transactions originating from that account must have authorized signatures from both Key Books. Managed transactions are introduced in Chapter \ref{managed}.

\subsection{Signature Threshold}

Each Key Page has a signature threshold. When a Key Page is used to authorize a transaction, whether as part of the origin account’s Main Key Book or Manager Key Book, the signature threshold determines how many signatures are required from the keys and/or Key Book Authorities of that Key Page. This is often referred to as $m-of-n$, where $m$ is the signature threshold, and $n$ is the number of entries. This general authorization scheme forms the basis for multi-signature support in Accumulate.

Signature thresholds are illustrated in Figure \ref{fig:key_m}, where \acs{ADI} `RedWagon’ submits a transaction through one of its Key Books and signs it using a number of keys that are specified in its priority 1 Key Page. In this $m-of-n$ authorization scheme, $n$ = 4 entries and $m$ is determined by the signature threshold. If the signatures are valid and the $m-of-n$ threshold is met, then the transaction is executed.

\section{Authorization}
Transactions are validated on most blockchains once a threshold number of signatures are submitted by accounts with access to the authorized private keys \cite{Bamakan2020}. Authorization schemes are defined by the number of signatures required to validate a transaction, with single signature and multi-signature schemes being the most common. As discussed in the previous chapter, Accumulate provides flexibility in key management and enables a user or group of users to add, remove, or modify the security or ownership of their keys through the use of hierarchical key sets. The ability to manage keys over time not only enhances multi-signature transactions but provides the means to create new authorization schemes that can replicate a variety of complex financial operations. This chapter describes how Accumulate processes single and multi-signature transactions and introduces several new authorization schemes that expand the utility of multi-party transactions.

\subsection{Single-signature Transactions}
Single signature transactions have an $m-of-n$ signature threshold of 1, meaning that one signature is required to validate a transaction. This authorization scheme is used by the majority of individuals to send and receive tokens due to its simplicity and speed. Most wallets automatically derive a signature from the user’s private key and transaction hash, so a transaction only requires a user to input the amount, destination address, and possibly a memo depending on the blockchain and the nature of the transaction. However, a single point of failure can result in a loss of funds, so users must be careful to protect their private keys.

When a user creates a transaction in Accumulate, they must specify the transaction type (e.g., data, tokens) and origin account. Each transaction specifies exactly one Key Page in the Key Book of that account, and all signatures must come from a Key on that page. If the transaction specifies Key Page 1, for example, and Key Page 1 specifies $m = 1$, then the transaction is immediately validated. This is similar to an embedded signature derived from a private key in a traditional blockchain address.

%ER: Move this part and the next paragraph to Key Management? 496 and 498
%A user may also specify $m = 1$ for several different Key Pages within their Key Book. This allows multiple parties to manage a single \acs{ADI} with single signature transactions single-signature transactions. This arrangement might be useful for distributing keys to minimize the damage to a compromised account or for organizing transaction types by Key Page.
%Rewrite as: Advanced features of Key Books allows you to do cool stuff with this, like a single Key Page with M = 1 and multiple keys so that anyone can sign any tx; however, they can also modify their own keys but will be able to lock key pages so you could set a flag to prevent that page from modifying itself. Single page with multiple keys and M = 1 without the danger of that page modifying itself. Many employees might not want them to be able to add keys or modify other employees’ keys, With M = 1, the page can modify itself, but we can lock the page to prevent that. Have Boolean on the page that says “locked” or not, so cannot modify keys. Paul suggests a “capabilities” field that determines that kind of txs the key page is allowed to do. Call it a flag for now. Can set or unset that flag to lock it. Can write this part vague here, but get into details in the Key Management chapter. Group of people with single sig, do not need to talk about the details.
%ER: Ask Paul: Should the threshold for key changes be different from that of normal transactions.

While the concept of Key Books, Key Pages, and Keys may appear complicated to the average user who simply uses their account to send and receive tokens, the authorization process from a user’s perspective will not look any different from a transaction executed on another blockchain when $m = 1$. The user will always have the option to upgrade or downgrade their security in the future without having to create a new account.

\subsection{Delegated Transactions} \label{delegated}
In addition to getting transactions approved by account managers within an \acs{ADI}, it could also be necessary for applications to disperse tokens based on external authorization(s). This is the basic mechanism of smart contracts and is important to many applications of blockchain technology. However, smart contracts have exhibited many inefficiencies, which arise from the fact that they are computationally expensive \cite{Chen2017}. Accumulate prevents this issue by limiting all transactions to a static amount of computation, ensuring that they take up a known amount of time, space, and energy on the blockchain. Within these constraints, Delegated Transactions allow for a system of third-party verification analogous to a smart contract while limiting their computational cost. Delegated Transactions allow for the transaction’s origin account and the signing Key Book to belong to different Root Identities. As the name suggests, this is done by delegating the signature authority of a key to another Key Book. The Key Book of the origin account contains the URL addresses of the external Key Books authorized to sign transactions against that account. A Delegated Transaction must be sent to the signer’s \acs{BVN} first, so that the signature can be authorized against the signer’s Key Book and added to the signer’s Signature Chain. It is then sent to the origin’s \acs{BVN} to be executed, at which point synthetic transactions are produced if necessary, for example, to disperse funds.
%REFERENCE for smart contracts being computationally expensive

%The Key Book of the Requester \acs{ADI} contains the URL addresses of the Key Books required to sign a transaction sent by a user of the \acs{ADI}. As previously mentioned, Accumulate does not allow submitted transactions to immediately change the state of the blockchain. Instead, the suggested transaction is added to a Signature chain within the local \acs{ADI} chain. A Delegated Transaction Request would then appear in the wallets of the signers. Upon signing the delegated transaction, synthetic transactions would be called to disperse the funds to the stipulated parties. 

%(Just a copy of what Ethan sent to Ben, no need to add) If the origin’s root identity (\acs{ADI}) and the signer’s root identity are different. To execute the transaction, the \acs{BVN} needs the origin’s state. To authorize the signature, the \acs{BVN} needs the signer’s state. Because we assume the root identities belong to different \acs{BVN}s, we must assume we cannot do both steps on the same \acs{BVN}.
%Therefore, there must be two \acs{BVN}s involved
%Therefore, there must be (at least) two transactions involved
%Part 1: On the signer’s \acs{BVN}, we authorize the signature; this is the delegated transaction
%Part 2: On the origin’s \acs{BVN}, we execute the user’s transaction; this is the inner transaction

The Key Book system allows for applications to tailor the number of required signers for a transaction to occur. For example, some transactions may only require the signature of a single authorized party, some require a $2/3$ majority, while some should require signatures of all involved parties. Each Key Page of a Key Book allows for a custom minimum number of signers to be applied to transactions. %Maybe insert a couple of sentences about the utility of delegated transactions for managing the funds of a foundation 

%(Possibly move to use cases) Another weakness of smart contracts that Accumulate aims to solve is the difficulties in changing their terms. Accumulate improves on this major shortcoming of smart contracts by making it easy to quickly remove a signer that has lost privileges without damaging the rest of the approval mechanism. The \acs{ADI} simply sends another transaction specifying the signer’s removal, which goes into effect once all authorized signers have approved this transaction. Failsafes can also be built into the Key Book to handle uncooperative signers. For example, a signer on a higher key page would be able to make changes to the list of signers on a lower ranked page.

%ER: add a paragraph or section about suggestions?

\subsection{Multi-signature Transactions}
Multi-signature transactions have an $n \geq m > 1$ signature requirement, meaning that the number of signatures required to authorize a transaction is less than or equal to the total number of authorized keys specified by the originator, but always more than 1. This authorization scheme is primarily used for adding extra security when transferring valuable assets or coordinating signatures between multiple accounts owned by individuals, groups, businesses, or automated programs with access to a key \cite{Xiao2021}. The main advantage of multi-signature transactions is the need for multiple points of failure in order for an account to be compromised. Several examples of the real-world utility of multi-signature authorization are provided below:

\begin{itemize}
    \item 1-of-2: A joint bank account where the signature of either member is sufficient to access the funds.
    \item2-of-2: Automated spending control in which a computer program with access to a key will only sign for a transaction below a certain dollar amount.
    \item2-of-3: Escrow transactions between two parties with a third party acting as an intermediary in the event of a dispute.
    \item1 or 3-of-4: Single sig with distributed account backup from trusted parties in the event the owner’s key is lost or stolen.
\end{itemize}

\begin{figure*}[]
 \centering
 \includegraphics[scale=0.70]{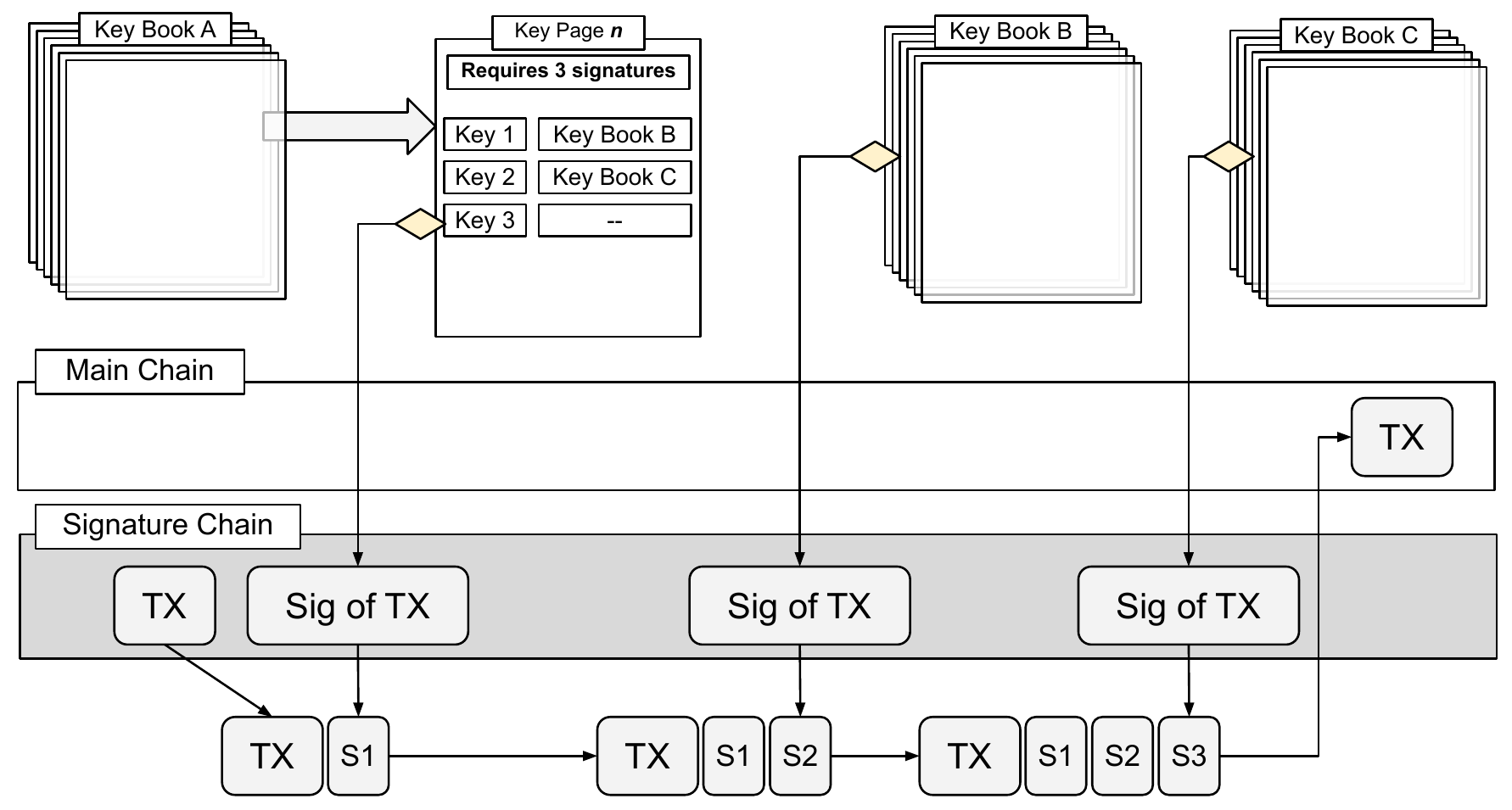}
 \caption{Overview of multi-signature transactions.}
 \label{fig:multisig}
\end{figure*}

All of the above scenarios are possible to model on the Accumulate network using multiple Key Pages and Delegated Transactions. For example, an individual \acs{ADI} may specify $n \geq m > 1$ for a single Key Page and control a multi-signature transaction entirely through their \acs{ADI}. If one or more authority Key Books from external \acs{ADI}s are specified in the Key Page of the origin account, then multiple \acs{ADI}s can participate in a multi-signature transaction through Delegated Transactions. This process is illustrated in Figure \ref{fig:multisig}.

Key Book A is the originator of a multi-signature transaction that requires at least 3 signatures in one of its Key Pages. Key Books B and C are specified as authority Key Books in the Key Page of Key Book A. The transaction is initiated by Key Book A, which signs with Key 3. An envelope is created from the hash of the signature and transaction and put onto the Signature Chain. Key Books B and C then sign the transaction with their keys, creating additional envelopes consisting of the hash of the original transaction and their respective keys. Once 3 signatures are collected on the Signature Chain, the transaction is promoted to the Main Chain. If the signature threshold is not met in approximately 2 weeks, then the transaction will be discarded. Signatures are discarded after 2 weeks regardless of whether the transaction succeeds.

\subsection{Managed Transactions} \label {managed}
\begin{figure*}[tb]
 \centering
 \includegraphics[scale=0.94]{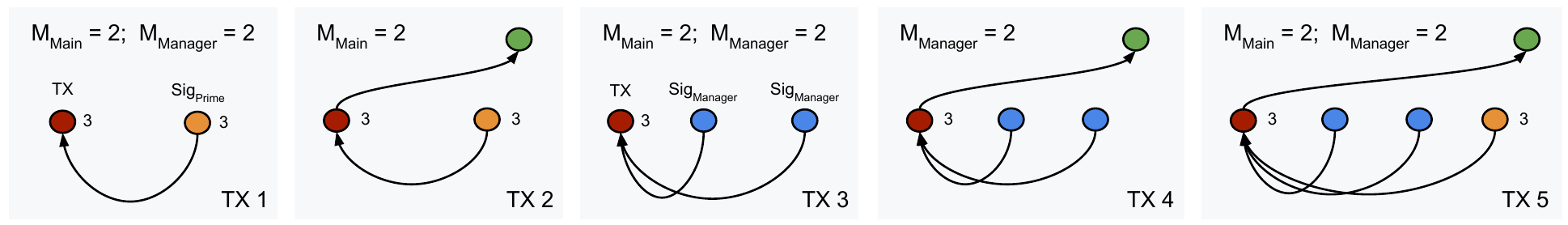}
 \caption{Managed transactions.}
 \label{fig:managed}
\end{figure*}

Every account has a Main Key Book and an optional Manager Key Book that users can specify when their account is created. The Main Key Book can initiate single and multi-signature transactions and specify multiple authority Key Books within a Key Page to participate in Delegated Transactions. A Manager Key Book uses its own $m-of-n$ authorization scheme to approve or reject any type of transaction initiated by the Main Key Book. While a manager can have multiple keys, an account can only specify a single manager. A manager is also unable to execute a transaction on their own. This makes managed transactions most useful for automated security without custodial control, as illustrated in the following examples.

\begin{itemize}
    \item Spending limits: A manager may enforce a spending limit similar to withdrawal limits on ATMs.
    \item Frequency limits: A manager may allow only a certain number of transactions per day.
    \item Transaction type limits: A manager may restrict transactions to certain types (e.g., tokens, data).
\end{itemize}

Transactions with various authorization schemes can be executed depending on whether a Manager Key Book, Main Key Book, or both Key Books are required to sign a transaction. Most users will specify both a Main and Manager Key Book to automate security for a user-generated transaction as an extra layer of security. For example, \acs{ADI} token account acc://alice/tokens may specify Key Book acc://bob/book as a manager. If a transaction is initiated by \acs{ADI} Alice, then a threshold number of signatures from the specified Key Page in \acs{ADI} Alice and any Key Page in \acs{ADI} Bob is required to validate the transaction. If a manager is specified, but $M = 0$ for a Main Key Book, then a manager can execute transactions that are initiated by an account. This authorization scheme may be useful for automating recurring payments (e.g., utility bills). If $M$ = 0 for both a Manager and a Main Key Book, or a Manager is not specified and $M$ = 0 for the Main Key Book, then anyone with the account information can execute a transaction. Due to the security risks involved, this authorization scheme may only be useful for data transactions among collaborators. These authorization schemes are formally defined below in terms of their $m-of-n$ signature requirements.

When a manager is not specified:
\begin{itemize}
    \item If $M = 0$, then anyone in the Accumulate network can execute transactions.
    \item If $M > 0$, then anyone can suggest a transaction, but at least one Key from the Key Page of the Main Key Book or from any Key Page in an authority Key Book also has to sign it.
\end{itemize}

When a manager is specified:
\begin{itemize}
    \item If $M = 0$, then anyone can suggest a transaction, but the manager also has to sign it.
    \item If $M > 0$, then anyone can suggest a transaction, but the manager and Key Page also have to sign it.
\end{itemize}

The process of signing managed transactions is illustrated in Figure \ref{fig:managed}. TX1 requires two Key Page signatures from Key Page 3 of its Main Key Book and two manager signatures from the Manager Key Book. While the Main Key Book signature requirement is satisfied due to one embedded signature and one independent signature, the transaction fails because there are no manager signatures. However, TX2 would be promoted to the Main Chain, represented by a green dot, because it does not specify a manager signature. TX3 has the same signature requirements as TX1. While two manager signatures have been provided, the transaction fails because there are no signatures from the Main Key Book. However, TX4 would be promoted to the Main Chain because it does not specify a signature from the Main Key Book. TX5 satisfies the signature requirements for both the Main and Manager Key Books, causing it to be promoted to the Main Chain.

\section{Architecture Overview} \label{architecture_overview}
The Accumulate protocol has a ‘chain of chains’ architecture in which account chains on a particular \acs{BVN} are treated as side chains of that \acs{BVN}’s root chain, and the root of each \acs{BVN} is treated as a side chain of the \acs{DN}’s root chain. This architecture enables massively parallel transactions and virtually limitless scalability as capacity is added in the form of additional \acs{BVN}s. While the reorganization of the blockchain as a ‘chain of chains’ built around accounts rather than blocks solves many persistent problems such as the scalability trilemma, it also introduces new challenges, including the efficient communication of receipts between \acs{BVN}s, the storage of data generated by accounts that function as continuously growing Merkle trees, and the coordination of consensus between multiple \acs{BVN}s and the \acs{DN}. This chapter provides a detailed overview of the account structure, data structure, data persistence, and system architecture that allows Accumulate to overcome these challenges and bring new innovations to the blockchain space.

\subsection{Data Structure}

\subsubsection{Stateful Merkle Trees}
A Merkle Tree is a balanced binary tree with a single root that can be derived from only a fraction of its hashes \cite{Dhumwad2017}. This allows for efficient verification that a particular hash and the transaction data associated with it is a genuine member of the tree. Accounts chains are continuously growing Merkle Trees that append, concatenate, and anchor new transaction hashes into their respective \acs{BVN} Root Anchor Chains every major and minor block. Due to the irregular frequency at which new hashes are added, accounts are expected to contain a variable number of roots within a block. This data structure is a special form of Merkle Tree, called Stateful Merkle Tree (\acs{SMT}). \acs{SMT}s may have multiple roots in different time frames and are constructed in a way that allows them to grow over time, whereas most Merkle tree variations are constructed once and never modified after the root is obtained. This is made possible by keeping a list of root hashes within a tree.
%We keep a list of root hashes so we can easily append new entries. Stateful cause we keep a list of root hashes to add to it, a normal Merkle tree you do not bother to maintain this list of roots, to maintain that state cause you do not grow it. Normally, you create a Merkle tree once, and it is done. Keep an array of hashes. 

%Typical Merkle Trees are not modified after they are created. To facilitate continuous growth, Accumulate retains enough state to 

%Accumulate uses Stateful Merkle Trees (SMTs)
%Typically, a Merkle Tree is not modified after it is created. 

%However, Merkle Trees in Accumulate grow continuously. In order to do this, we must retain enough state. 

\begin{figure*}[t]
 \centering
 \includegraphics[scale=0.85]{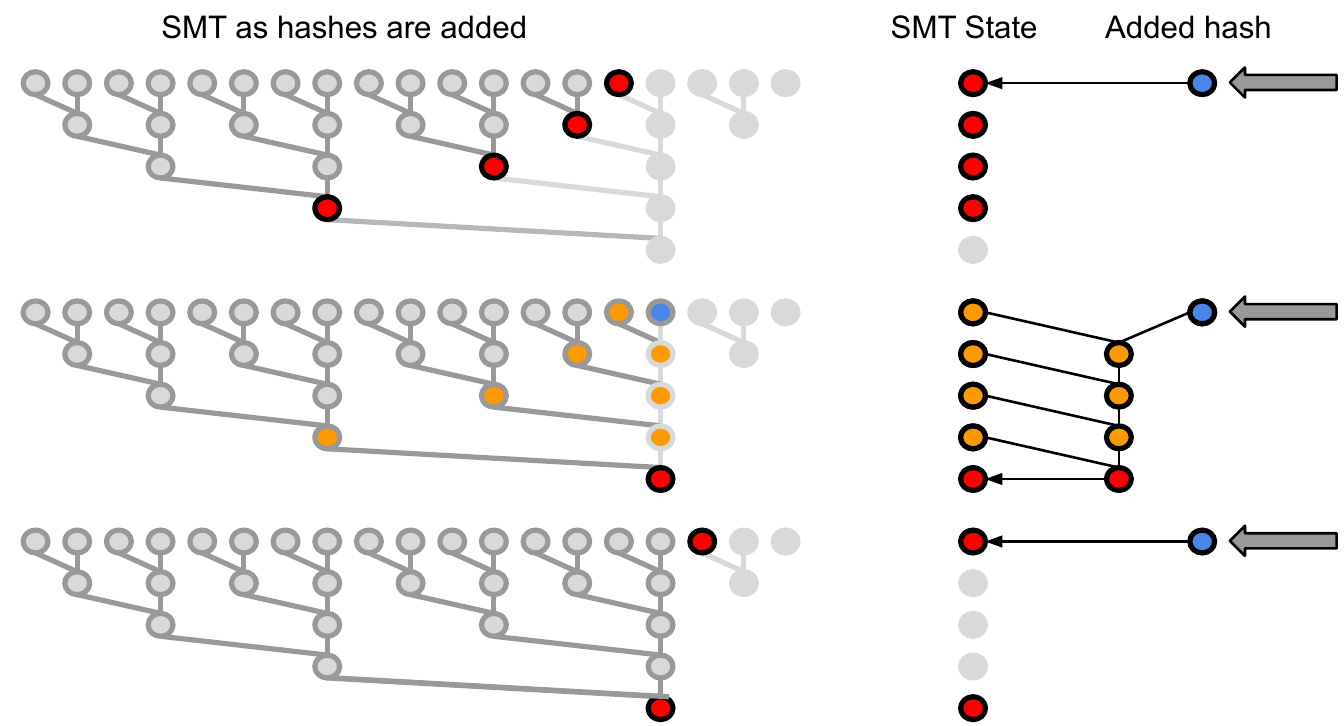}
 \caption{Adding hashes to a Stateful Merkle Tree.}
 \label{fig:data}
\end{figure*}

\acs{SMT}s provide the support for building and maintaining information in accounts across multiple blocks. An \acs{SMT} can collect hashes of validated data from multiple sources and create a Merkle Tree that orders all these hashes by arrival time to the \acs{SMT}. The arrival order of entries from all sources is maintained in the tree. \acs{SMT}s can also be fed into other \acs{SMT}s to create the superstructure of the Accumulate network in which all accounts that are managed by a particular \acs{BVN} are anchored into its respective Root Anchor Chain, and all \acs{BVN} Root Anchor chains are anchored into the \acs{DN}. 

An \acs{SMT} is append-only, meaning that hashes are added from left to right, and a concatenated hash pair (i.e., parent) is added as soon as two children hashes exist on the same level of the tree. Unlike a traditional Merkle tree, an \acs{SMT} has multiple roots. Therefore, we need a method to compute a single root from all the roots of every major and minor block. The process of combining roots is illustrated in Figure \ref{fig:data}, in which new transaction hashes are appended to a growing \acs{SMT}.
%Changed Figure 8.1 to reference Figure 7
In this diagram, blue dots represent transaction hashes, red dots represent the root of a particular subtree that defines its current state, and orange dots represent the previous roots that defined the previous state. The \acs{SMT} is constructed on the left, while the column on the right summarizes the current state at each level where Level 1 is considered the entry point of a new transaction hash at the leaves of the tree.

We begin 15 transaction hashes into the \acs{SMT}, where 4 different root hashes in 4 independent subtrees summarize the current state of the tree. When the next hash is added, the system recognizes that a root hash already exists on Level 1 to the left of the incoming hash. The two children are automatically concatenated to produce a root hash at Level 2. However, the system also recognizes that a root hash already exists on Level 2 to the left of the parent root hash. This process continues until the root hash on Level 4 is concatenated with the parent hash created on Level 4 to produce a final root hash on Level 5. Note that the current state at this point can be represented by a single hash. When the next hash is added on Level 1, it is treated as a second root hash since no other roots exist to the left.

To calculate an anchor at any point in time, the rightmost hash is successively concatenated with the root hash to the left until a single hash remains. For example, the four root hashes in Figure \ref{fig:data} can be concatenated into an anchor without the addition of a new transaction. Since the entire state of the \acs{SMT} is captured in the root hashes, the previous hashes can be discarded or `pruned' from the tree to minimize its size. The entry hashes are saved in regular accounts on the Main Chain to maintain a historical record of each entry. However, entry hashes older than 2 weeks are pruned from Signature Chains such that only the root hashes of the current state are recorded. The process of pruning entry and intermediate hashes on the Signature Chain forms the basis for Scratch Accounts, which are introduced in Chapter \ref{scratch_chains}.

\subsubsection{Binary Patricia Tries}

Patricia Tries \cite{Jung2002} are generally designed to create small cryptographic proofs about the particular state of values at a particular point in time. On a blockchain, a Patricia Trie can be used to prove the balance of an account (i.e., an address) at a particular block height. Since Accumulate organizes the blockchain into a series of chains under the management of different accounts, we need the ability to prove the state of a chain at a particular block height and index the block where a chain was last modified. The Binary Patricia Trie (\acs{BPT}) provides these proofs for the state of each chain.

A \acs{BPT} is similar to a Merkle Tree in that both hash a set of values that are ultimately concatenated into a single root. However, the \acs{BPT} also adds a key per entry to create a key/value pair where the key refers to the hash of a chain's URL (e.g., acc://RedWagon). While a Merkle Tree organizes entries by the order in which they are added, a Patricia tree organizes records according to the key of each record. The major differentiating feature of a \acs{BPT} is that the order in which entries are added does not matter. Within Accumulate, the \acs{BPT} serves as a cryptographic summary of the current state of every chain and account within a \acs{BVN}. This enables the network to reach consensus at a frequency of one minor block and guarantee that all nodes have the same state. Additionally, the \acs{BPT} is used to create snapshots of the state of the network, which new nodes can use to get up to date instead of executing all of the transactions.

\subsubsection{\acs{BPT} Implementation}
The Accumulate protocol has three entry types in the \acs{BPT}:

\begin{itemize}
    \item Node. Node entries are used to organize the tree. They have a left path and a right path and exist at a height in the \acs{BPT}. The height is used to consider a particular bit in the \acs{BPT}. The Left path is taken if the key has a zero bit at that point. The Right path is taken if the key has a 1 at that point.
    \item Value. The key-value pair in the \acs{BPT}. Value entries have no children, and paths through the \acs{BPT} from parent to child Node entries must end with either a nil or Value entry.
    \item Load. Node represents the fact that the next Byte Block is not loaded and needs to be loaded if the search is running through this part of the \acs{BPT}.
\end{itemize}

The most important feature of a \acs{BPT} is its ability to quickly update its state to provide proof of the state of the entire protocol. This ensures that changing the order in which entries are added to the \acs{BPT} has no effect on the result. The same set of entries will always produce the same structure. Further, subsets of the \acs{BPT} are independently and provably correct so long as there is a path from the root of the Patricia Tree to one of its subsets.

Many implementations of Patricia Trees are described in the literature \cite{Szpankowski1990}. In the Accumulate protocol, keys are randomly distributed from a binary point of view because the keys are the hashes of URLs. Mining these hashes to build some particular pattern of leading bits is not very manageable or possible as the keys are derived from URLs. URLs can be mined to have interesting leading bits, but little incentive exists to do so.
%MAYBE ANOTHER REFERENCE TO THE FIRST SENTENCE ABOUT MANY IMPLEMENTATIONS THAT ARE FOUND IN THE LITERATURE
Given that the keys created from hashes can be relied on to be numerically random, a \acs{BPT} will be very well-balanced if this feature is exploited. Note that we use the leading bits to organize entries in the \acs{BPT}. However, nothing prevents us from using every 3rd bit, the trailing bits, or any other random walk of bits in the key. Should an attack be mounted to create chain IDs that significantly unbalance the \acs{BPT}, we can refactor the \acs{BPT} using any of these methods and do so over time by reorganizing only parts of the \acs{BPT}.

\subsubsection{Summary}
The organization, membership, and state of chains with an account can be summarized as follows:

\begin{itemize}
    \item Accounts can act as domains to allow addressing of their Main and Signature Chains as a set of URLs.
    \item The membership of Chains is proven using Stateful Merkle Trees.
    \item All entries in a chain are members of one continuously growing Merkle Tree.
    \item At any point in time, the membership of the Merkle Tree is provable by the Merkle State.
    \item \acs{BPT}s can prove the state of every account and chain in a \acs{BVN} at a particular block height.
\end{itemize}

The Accumulate protocol is able to maintain proof of its state without requiring all elements of that state to be at hand. This allows the network to efficiently scale and makes it possible for mobile devices to prove the entire state of accounts relevant to the user. Accumulate organizes these proofs as a very large set of chains. Even token transactions are organized as chains where the membership and order of transactions from a particular account exist in its own chain.

Within Accumulate, the \acs{BPT} captures account proofs in the \acs{BPT} as key/value pairs, where the `key' is a hash derived from the account URL, and the `value' is the final hash of a proof. Account proofs are constructed as transient Merkle Trees that take as inputs the hash of the account state and an anchor from each of the account's chains. The summary of all key/value pairs is a hash that acts like the Merkle Root of a Merkle Tree. Additions to or modifications of this Merkle Tree-like structure only require localized re-computations, making it easy to update the state of the entire network at the frequency of a minor block.

\begin{figure*}[t!]
 \centering
 \includegraphics[scale=0.85]{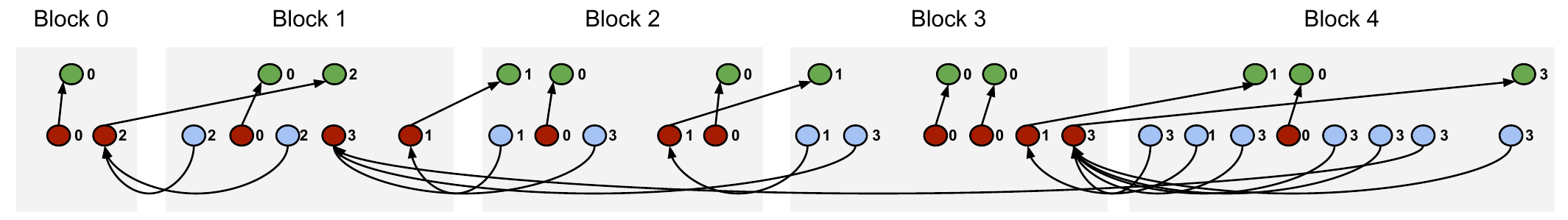}
 \caption{Collecting signatures on the Signature Chain and promoting the transaction hashes to the Main Chain.}
 \label{fig:pending}
\end{figure*}

\subsection{Account Chains}
Each account has a separate Signature Chain and Main Chain. The Signature Chain is responsible for collecting signatures, while the Main Chain is responsible for executing transactions that have met the signature threshold.

\subsubsection{Signature Chain}
All transactions submitted to the Accumulate network are added to an account's Signature Chain. This includes data and token transactions submitted by Lite Accounts, data and token transactions submitted by \acs{ADI}s, managed transactions approved by a Manager Key Book, and Delegated Transactions submitted by an external Key Book. Single signature transactions are immediately promoted to the Main Chain if the signature included in the envelope is valid and corresponds to a key in the origin account's Key Page. Multi-signature transactions are not promoted to the Main Chain until the $m-of-n$ signature threshold is reached.

The primary utility of the Signature Chain is the ability to submit signatures independently of transactions. This is accomplished by hashing the transaction once by itself and once again as an envelope that includes the hash of the transaction and the hash of the signatures. Adding the envelope hash to the Signature chain provides cryptographic proof that the signatures were obtained. The first person to submit the multi-signature transaction sends the raw transaction along with their signature to create the first envelope. Once the signature threshold is reached, the transaction is promoted to the Main Chain.

Signatures are collected completely on-chain and over a reasonably long time frame so that signors can be notified, transactions can be modified, and signatures can be collected. Moving these activities on-chain provides an audit trail of the consensus-building process. The lifetime of signatures and transactions submitted to the Signature Chain is approximately 2 weeks. This means that a transaction will fail if the $m-of-n$ signature threshold is not reached within the allotted time.

\subsubsection{Main Chain}\label{main_chain}
The Main Chain of an account permanently stores transaction hashes without their signatures because the signatures have already been validated on the Signature chain. The data structure of the Main Chain is similar to that of the Signature chain in that each account's Main Chain is a continuously growing Merkle Tree. After each major block, any entries on the Signature Chain older than 2 weeks are discarded.

The illustration provided in Figure \ref{fig:pending} shows the process of collecting signatures on the Signature Chain and promoting the transaction hashes to the Main Chain. The top row represents the Main Chain, while the bottom row represents the Signature Chain. Blue circles represent signatures, and red circles represent envelopes for four Key Pages within a Key Book where:

\begin{itemize}
    \item Priority 0 specifies 1 key required.
    \item Priority 1 specifies 2 keys required.
    \item Priority 2 specifies 3 keys required.
    \item Priority 3 specifies 6 keys required.
\end{itemize}

The transaction managed by Priority 0 Key Page only requires a single signature to satisfy the $m-of-n$ requirement. If a transaction is suggested in a previous block, but the final signature is provided in a later block, the transaction is promoted to the Main Chain in the latter block. For example, the transaction managed by the Priority 2 Key Page in Block 0 and signed by the second key in Block 1 is also submitted to the Main Chain in Block 1. Signatures must also be provided from the appropriate Key Page. For example, a signature from the Priority 2 Key Page cannot sign a transaction from the Priority 3 Key Page.

\subsection{Scratch Accounts}\label{scratch_chains}
The architecture of Accumulate allows for the efficiency of the state in validating transactions, collecting signatures, and coordinating consensus. Transactions can be compressed into 12-hour blocks because the Accumulate blockchain is a collection of Merkle trees with arbitrary sync points. Mobile devices can be used as Lite nodes to validate only those transactions that are relevant to an account because each account is treated as an independent blockchain. Signatures can be pruned before a multi-signature transaction is validated because Accumulate has a transient Signature Chain and a permanent Main Chain. These features can also be combined to create Scratch Accounts, a specialized type of account with limited availability that facilitates cheap on-chain consensus between multiple parties and provides cryptographic proof of validation without overburdening the blockchain.

\subsubsection{Implementation of Scratch Accounts}
\acs{ADI} Data Accounts and \acs{ADI} Token Accounts can be marked as Scratch Data or Scratch Token Accounts when the account is first created. The only difference between an \acs{ADI} account and a Scratch Account is that the data availability for a Scratch Account expires after approximately 2 weeks, which is similar to the lifetime of transactions and signatures present on the Signature Chain. Once the data expires, the account is compressed, and a proof of the transactions is created. This proof contains less data than the Scratch Account it represents, which allows the blockchain not to be burdened with moving substantial amounts of data over the main network.

To view the current state of data or token transactions in a Scratch Account, a user can query the latest entry on the chain using the following API commands:\\

\noindent $Query(acc://adi/account\#data))$\\
$Query(acc://adi/account\#transaction))$\\

A user can also query the history of data or token transactions in a Scratch Account over a period of time by specifying an arbitrary range:\\

\noindent $Query(acc://adi/account\#data/0:10))$\\
$Query(acc://adi/account\#transaction/10:50))$\\

The main parameter of $query$ is a URL, which contains generic labels for $adi$ and $account$ that are specified by the user. The end of the URL is appended with the $\#data$ or $\# transaction$ fragment, which directs the search to the appropriate account type. The range (e.g., 0:10) provides a historical record of entries in sequential order. A query of 0:10 would retrieve records beginning with entry 0 and ending with entry 9. However, any entries older than 2 weeks will be pruned, meaning that they will not be stored by the protocol.

\subsubsection{Pruning Data}
Pruning is a data compression strategy designed to reduce the size of a data tree by removing unnecessary or redundant data \cite{Palm2018}. In a typical blockchain, pruning refers to the removal of intermediate and leaf (transaction) hashes once the transaction outputs on which those hashes rely have been spent. This strategy significantly reduces the data footprint of the blockchain and allows a pruned node to fulfill many of the same functions as a full node. On Accumulate, pruning applies to Signature Chains and Scratch Accounts. All signatures on a Signature Chain and all transactions on a Scratch Account are automatically pruned after approximately 2 weeks.

Unless an account is deliberately purged by a user, the Merkle tree will still contain the historical record of a Scratch Account after its data is pruned. Transaction hashes and intermediate hashes are deleted, but the root hashes that summarize the state of a Scratch Account will remain. This is illustrated in Figure 8, where a Stateful Merkle Tree is being created from its transaction hashes. In a typical account, transaction hashes in the top row of the Merkle tree are stored by the data servers to maintain a record of every entry. Root hashes (red) that define the current state of the account are also stored by the data servers. In a Scratch Account, transaction hashes are pruned, but the root hashes are saved on-chain. This allows any user with access to the transaction hashes to cryptographically prove the data in a Scratch Account by deriving the root.

\subsubsection{Utility of Scratch Accounts}
Scratch Accounts allow a user or organization to build arbitrarily complex use cases for a low cost without burdening the blockchain. The following list provides several real-world applications that can benefit from Scratch Accounts.

\begin{itemize}
    \item Isolating events within high-frequency data: \acs{IoT} networks can collect high-frequency data on Scratch Accounts, prune baseline data, and promote important events (e.g., temperature fluctuations) to the Main chain. Baseline data can be periodically compressed to produce summary information about a system.
    \item Deriving oracle data: Miners can scrape pricing information from public pricing feeds, mine the pricing data, and submit records on Scratch Accounts for rewards at a fraction of the cost of writing data to the Main Chain. The winning data is put on the Main Chain while the losing data is pruned.
    \item Proving transactions on exchanges: It is cost-prohibitive for a major exchange to provide cryptographic proof of every transaction. With Scratch Accounts, exchanges could store transaction data off-chain and cheaply derive root hashes from pruned data.
\end{itemize}

\subsection{Anchoring}
Anchoring is the process of inserting external data into a blockchain transaction to provide an immutable receipt that can be verified by anyone with access to the original data \cite{Wang2021}. This is typically accomplished by hashing the data on or off-chain, constructing a Merkle tree from that data, and periodically inserting its Merkle root into a larger and more secure blockchain. Factom, the predecessor of Accumulate, was one of the first protocols to anchor into Bitcoin and Ethereum. By hashing its transactions into larger directory blocks and combining directory blocks into larger anchors, Factom was able to maintain an immutable record of thousands of hashes at the cost of a single transaction.

Accumulate retains the ability to anchor into other protocols, but it also requires internal anchors as a consequence of its chain-of-chains architecture. Accounts are connected by anchoring their roots to their respective \acs{BVN}s, \acs{BVN}s are connected by anchoring their roots to the \acs{DN}, and blocks are connected by anchoring the root of the \acs{DN} into each \acs{BVN}. Anchoring each of these elements allows Accumulate to be a blockchain network as opposed to a collection of independent protocols.

\begin{figure*}[th]
 \centering
 \includegraphics[scale=0.8]{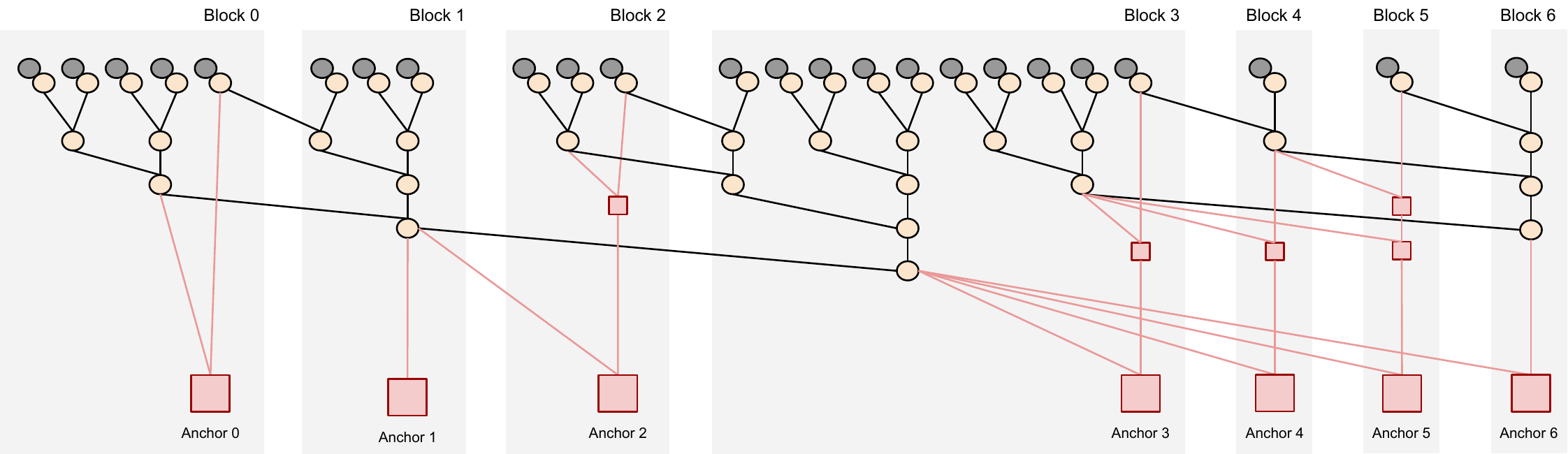}
 \caption{Anchoring transactions within a Stateful Merkle Tree.}
 \label{fig:anchoring}
\end{figure*}

\subsubsection{Anchoring Within a Chain}
In most blockchains, the beginning and end of a Merkle tree are bounded by the block time. In Accumulate, the end of a block is treated as synchronization points that simply designate when transactions should be anchored. Merkle trees transcend blocks, which means that every chain in Accumulate can be modeled as a continuously growing Merkle tree. However, the freedom to append new hashes on a continuous basis may result in a Merkle tree having multiple roots at any point in time. This bloats the network and adds unnecessary complexity to the protocol when validating transactions. 

Accumulate addresses this issue by organizing its hashes with a Stateful Merkle Tree (\acs{SMT}). Hashes from any account chain (e.g., Main Chain) or system chain (e.g., Synthetic Transaction Chain) that changed since the previous block are concatenated into roots at the end of the current block. Multiple roots are concatenated into a single root (i.e., anchor). This process is illustrated in Figure \ref{fig:anchoring} for a generic account chain where gray rectangles represent blocks, gray circles represent raw transactions, yellow circles represent hashes, and red squares represent anchors. For the sake of clarity, we refer to each row of hashes as a "Level," where Level 0 refers to the top row (i.e., leaf) hash of a raw transaction.

At the end of every block, hashes are concatenated into roots. In Block 0, for example, one root is produced from transactions 1-4, and a second root is produced from transaction 5. These roots are concatenated into a single anchor labeled "Anchor 0". If the root of a previous block is on a higher level than the root of the current block, then these roots must be connected through intermediate anchors before a final anchor is produced. Block 2, for example, has two root hashes on Levels 0 and 1. These are concatenated into an intermediate anchor on Level 2. This intermediate anchor is concatenated with the root on Level 3, Block 1, to produce Anchor 2. This anchor summarizes the entire history of the account chain until this point because it contains every hash from Blocks 0-2. 

The number of intermediate anchors required to produce a final anchor that summarizes the history of the chain until that point is equal to the number of roots in previous blocks that are at a lower level than the root in the current block. Block 5, for example, has a single root at Level 0. Block 4 has a single root at Level 1. Block 3 has two roots at Levels 2 and 4 that are higher than Level 0. Therefore, Block 5 needs two intermediate anchors to produce Anchor 5.

\subsubsection{Anchoring Within a \acs{BVN} and the \acs{DN}}
Every chain in a \acs{BVN}, except for the Binary Patricia Trie (\acs{BPT}), appends its hashes to an \acs{SMT} whose anchor represents the current history of its chain. Every chain, with the exception of the Root Anchor Chain, inserts its key/value pairs into the \acs{BPT}, whose anchor represents the current state of its chain. These anchors are then fed into the Root Anchor Chain, whose anchor represents both the history and the state of every chain within a particular \acs{BVN}. This process occurs in parallel within every \acs{BVN} in the network and also within the \acs{DN}, which contains its own \acs{BPT}, Synthetic Transaction Chain, and a Main and Signature Chain for the ACME token issuer.

The process of collecting hashes, producing roots, and anchoring roots into other chains can be modeled as a fractal. This is illustrated in Figure \ref{fig:chains} for the \acs{DN}, an arbitrary number of \acs{BVN}s, and an arbitrary number of account chains within a \acs{BVN}. This hierarchical organization of hashes is the basis for calling the Accumulate network a chain-of-chains.
%This was the proper figure, but I made sure it tracked to Figure 2

\subsubsection{Anchoring between a \acs{BVN} and the \acs{DN}}
Accumulate partitions its network through the use of \acs{BVN}s, which are independent networks of Tendermint nodes that process transactions in parallel. Scaling is achieved by adding more \acs{BVN}s. However, partitioning can also lead to inefficiencies in communication between \acs{BVN}s. Consider the case of 3 \acs{BVN}s in the absence of synthetic transactions or the \acs{DN}. An account on \acs{BVN}-1 sends a transaction to an account on \acs{BVN}-2, but neither \acs{BVN} has a complete record of the account balances of the other so queries need to be made to all 3 BVNs. Mathematically, this is represented by $N*N-1$ total queries. As the number of \acs{BVN}s increases, the number of queries required to communicate debits and credits will grow exponentially.

Accumulate addresses this issue by sending anchors and synthetic transactions from each \acs{BVN} to the \acs{DN}, then sending an anchor from the \acs{DN} to each \acs{BVN}. This process can be visualized by once again considering a network of 3 \acs{BVN}s. An account on \acs{BVN}-1 sends a transaction to an account on \acs{BVN}-2, which results in the creation of a synthetic transaction by \acs{BVN}-1. In order for BVN-2 to trust the synthetic transaction, it must be able to prove that the synthetic transaction was included in a completed DN block. This synthetic transaction is concatenated with other synthetic transactions generated by accounts in \acs{BVN}-1 that block and sent as an anchor to the \acs{DN}'s Intermediate Anchor Chain. The intermediate Anchor Chain collects synthetic transaction anchors from \acs{BVN}s 1, 2, and 3 and combines them into a single anchor within its Root Anchor Chain. This anchor is sent back to \acs{BVN}s 1, 2, and 3 as a receipt. \acs{BVN}-1 validates the anchor and creates a receipt for the synthetic transaction, starting with the hash of the synthetic transaction and ending with an anchor from the \acs{DN}'s Root Anchor Chain. It packages up that receipt as a signature and sends this as a synthetic transaction to \acs{BVN}-2. However, \acs{BVN}-2 has also validated this block of the \acs{DN}, so a simple Merkle proof is all that's needed for \acs{BVN}-2 to prove this synthetic transaction. This process reduces the number of queries to $2N$, which allows the network to scale linearly as more \acs{BVN}s are added.

\subsubsection{The Utility of Anchoring}

Bitcoin was originally designed as a trustless and permissionless alternative to cash that was free from the control of centralized banks and third-party intermediaries. As businesses began to invest in Bitcoin and offer financial products such as Bitcoin ETFs, Bitcoin evolved from a peer-to-peer network for sending digital currency between individuals to a store of value and a hedge against inflation. However, the institutional adoption of Bitcoin for purposes other than investment has been relatively slow to develop. Businesses have little incentive to adopt a trustless solution if they have established relationships with partners that they already trust, and regulation limits the activities that a user can engage in without permission.

Accumulate believes that the business value of Bitcoin lies primarily in its utility as an immutable historical ledger for anchoring external data. This concept gained mainstream attention in 2015 when DocuSign partnered with Visa in a pilot study to anchor DocuSign contracts into Bitcoin transactions using Visa's payment rails. External data can also be anchored into Bitcoin by other blockchain protocols. Factom, for example, regularly anchored its directory blocks into Bitcoin and Ethereum to bolster the credibility of its Proof of Authority (PoA) consensus mechanism. 

Accumulate expands the utility of anchoring through its use of internal and external anchors by virtue of its multi-chain architecture and cross-chain capabilities. Internal anchors can be used for cryptographic timestamping within the Accumulate network. External anchors can be exchanged with any blockchain protocol that produces a summary hash, allowing Accumulate to act as a universal bridge between the blockchain and the business world. This bridge enables a variety of business use cases, a few of which are summarized below:

\begin{itemize}
    \item Tamper-proof records: Accumulate will regularly anchor its \acs{DN} root hash into Bitcoin transactions to create timestamped receipts. Any customer with access to the original data can derive the anchor and cryptographically prove the validity of the receipt. 
    \item Proof of existence: Internal anchors are compact historical records that can be used to prove the existence and order of transactions at the frequency of a minor block. The valuation of a piece of real estate at a given point in time, the purchase of an insurance policy before a claim is submitted, and the use of a license before a contract expires can all be managed within the Accumulate network.
    \item Cross-chain bridges: Anchoring allows Accumulate to extend its Merkle proof into another blockchain. For example, data from Tezos can be validated through an Ethereum smart contract using Accumulate as a bridge.
\end{itemize}

%\subsubsection{Take Layer-1 anchoring stuff from Factom whitepaper and website. See FAQ. Anchoring the \acs{DN} to Layer-1 Blockchains}

\begin{figure*}[tb]
 \centering
 \includegraphics[scale=0.85]{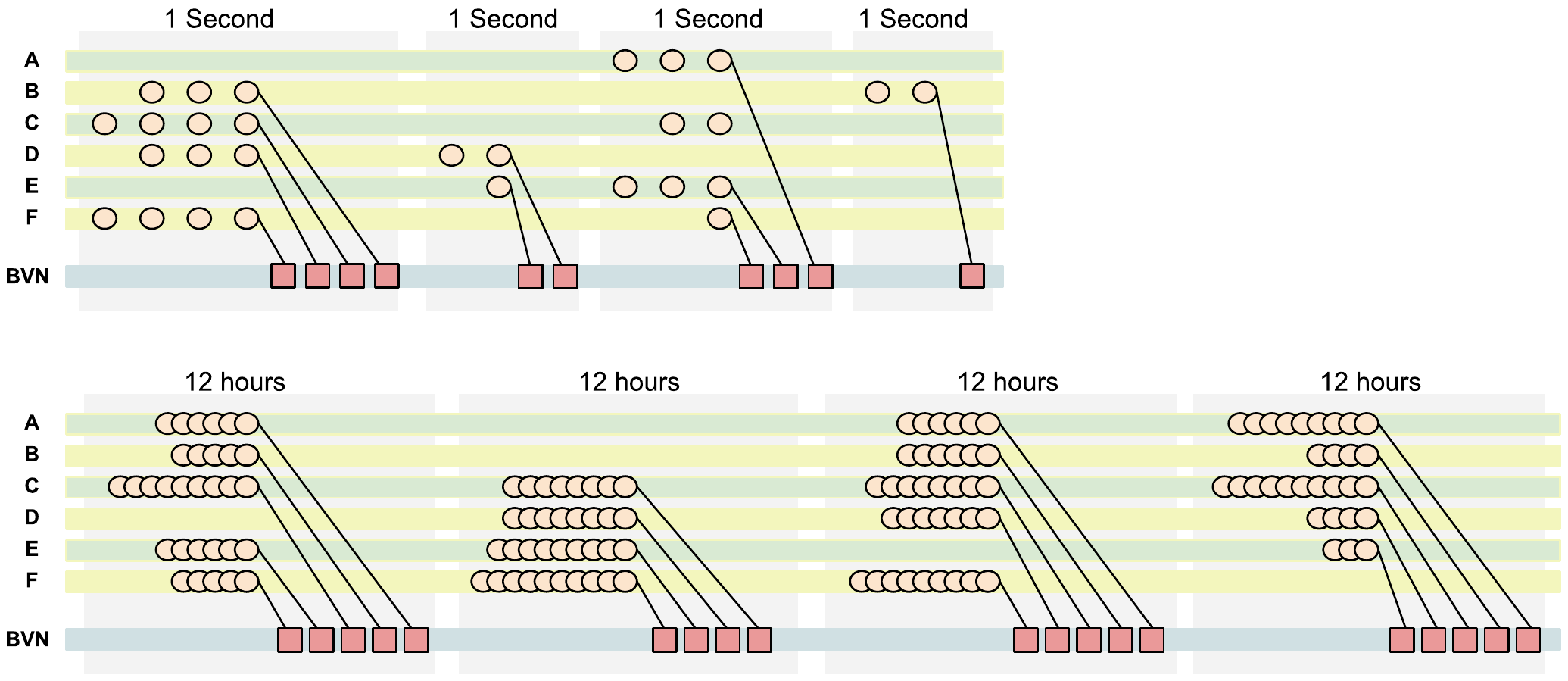}
 \caption{Major and minor blocks.}
 \label{fig:major}
\end{figure*}

\subsection{Major and Minor Blocks}\label{major_minor}

\subsubsection{Organization of Blocks}
In a traditional blockchain, all transactions submitted by all accounts within a block are concatenated into a root hash inside of a single Merkle tree \cite{Gamage2020}. A new Merkle tree is generated in the subsequent block and tied to the state of the previous block by including the previous root hash in its header. Synchronization between nodes on the network is handled on a per-block basis, which means that transactions cannot be processed until a new block is created. This typically occurs on the scale of minutes for Proof of Work (\acs{PoW}) blockchains like Bitcoin \cite{Bach2018} and seconds for Proof of Stake (\acs{PoS}) blockchains like Algorand and Ethereum 2.0.

The Accumulate protocol takes a different approach by organizing itself around accounts that define blocks rather than blocks that contain Merkle trees. Each account is a continuously growing Merkle tree with arbitrary blocks that can be validated at multiple synchronization points to coordinate consensus between \acs{BVN}s. These synchronization points are defined as Minor and Major blocks, with approximate block times of 1 second and 12 hours, respectively. 

\subsubsection{Minor Major Block Architecture}

Minor blocks are used to establish what set of accounts within a particular \acs{BVN} changed within the last second. Major Blocks use Minor Blocks to establish what set of accounts changed in the last 12 hours. Changes to the Transaction and Signature Chains of each account and to the Synthetic Transaction Chain of a \acs{BVN} are recorded as root hashes that are anchored into the Minor and Major Anchor Chains of that particular \acs{BVN}. This process is illustrated in Figure \ref{fig:major} for hypothetical accounts A-F across several minor and major blocks on a single \acs{BVN}.

Yellow circles represent transaction hashes submitted by accounts A-F. These transaction hashes are concatenated in the Merkle tree of each account until a root hash is derived. Red squares represent the root hashes that are anchored to the \acs{BVN} Minor and Major Anchor Chains. The order in which each anchor is added to the Anchor Chain is determined by the account ID, which is derived by hashing the entire account URL and collecting the first 8 bytes of the hash.

\subsubsection{Benefits of Multiple Block Times}

Most blockchains are committed to a single block time because each block is a part of the entire network with a well-defined beginning and end. However, there is no block that represents the entire Accumulate network because each account is treated as an independent blockchain. Similarly, transactions submitted by an account are processed by one of many \acs{BVN}s in the network that coordinate consensus but otherwise operate independently. This architecture allows Accumulate to define multiple block times of arbitrary length without disrupting the network.

The use of multiple block times allows the Accumulate network to quickly process transactions at speeds comparable to the fastest \acs{PoS} protocols while giving users the ability to store the entire state of the network on a mobile device. Indexing on two different time scales allows for the possibility of reducing the required storage space by increasing the granularity of indexing for past transactions. For example, entries on the minor anchor chains may be discarded after 2 weeks to save space. As a result, the granularity of the data is reduced to 12 hours after that point.

\subsection{Synthetic Transactions}

\subsubsection{Database Scaling}

How a business chooses to store and process its customer data largely depends on the size of the database, its read/write volume, and the availability requirements of its data. A small business may find it more convenient and economical to store its data in a database handled by a single server and vertically scale its hardware by upgrading its RAM, CPU, and storage capacity. It may also choose to replicate its database across multiple servers to make its data available even if some hardware fails or a server crashes. However, vertical scaling is not a viable option for a larger organization like Facebook, which generates upwards of 4 petabytes of data per day.

Instead, large and data-centric corporations often use a horizontal scaling strategy called sharding that distributes a single dataset across multiple databases and stores these databases on multiple servers or nodes. With this architecture, scalability and throughput are no longer limited by hardware but by capacity. Adding capacity by installing additional servers allows a database to scale while distributing the data across different shards increases throughput by allowing parallel read/write operations.

\subsubsection{Scaling Solutions in Blockchain}
Blockchain also utilizes a form of horizontal scaling, which relies on a system of rewards for those who do the work of validating transactions and securing the network \cite{Zhou2020}. Stakers in \acs{PoS} blockchains and miners in \acs{PoW} blockchains lend their tokens or computational power to confirm transactions. Full node operators for \acs{PoS} blockchains store copies of the entire blockchain and secure the network, receiving rewards at the risk of losing their stake if they engage in bad behavior.

A major difference in database management between traditional enterprises and blockchain protocols is the decentralization of the latter. Unfortunately, this also introduces the Scalability Trilemma, which can be defined as a trade-off between decentralization, security, and scalability. Bitcoin, for example, chose security over scalability when it introduced a delay in the form of blocks. EOS, meanwhile, chose scalability at the cost of decentralization.

\subsubsection{Scaling Solutions in Accumulate}
Accumulate bypasses the scalability trilemma by organizing itself around digital identities (e.g., \acs{ADI}s, Lite Accounts) and treating each account as an independent blockchain. All accounts and the data contained within can be operated upon independently by design and validated by a series of \acs{BVN}s that operate in parallel.

Each \acs{BVN} is a network of Tendermint nodes that is responsible for validating transactions initiated by an account. Each account is assigned to a particular \acs{BVN}, and as more identities are created, more \acs{BVN}s can be added to scale the network and maintain high throughput. The summary hashes of an account's current state are anchored to the \acs{BVN}. Meanwhile, each \acs{BVN} is anchored to the \acs{DN}, which is responsible for coordinating the \acs{BVN}s and guaranteeing security for the network. Security is further enhanced by anchoring the \acs{DN} to a secondary blockchain such as Ethereum or Bitcoin to gain the added security provided by those networks.

While each identity acts as its own independent blockchain, transactions between the \acs{BVN}s that host these identities necessitate a query of the sender's transaction history to determine if they actually possess the assets they wish to send. Since no \acs{BVN} contains the entire state of the Accumulate network, each \acs{BVN} must be independently queried. As more \acs{BVN}s are added to scale the network, the majority of queries will involve \acs{BVN}s that have no transaction history with the sender. This puts an unnecessary load on each \acs{BVN}, which increases transaction cost and decreases transaction throughput.

To solve this problem, a third type of node network is introduced, which is called the Data Server Network (\acs{DSN}). The \acs{DSN} does not have the requirement of high validation throughput. It will move the transaction data from the \acs{BVN}s and organize the transaction history into organized chains. This reduces the resource load on the \acs{BVN} nodes, enabling the \acs{BVN}s to focus on managing the identity states while allowing the \acs{DSN} to maintain the transaction history for those states.

\subsubsection{Introduction to Synthetic Transactions}
To optimize transaction throughput, Accumulate uses synthetic transactions, which can be broadly defined as any transaction that is produced by the protocol rather than by the user \cite{Michelson2021}. To understand how synthetic transactions work, consider the classic example of Alice, Bob, and Charlie, who will be treated as separate \acs{ADI}s on the Accumulate network.

Alice's transactions are handled by \acs{BVN}-1, Bob's by \acs{BVN}-2, and Charlie's by \acs{BVN}-3. Bob needs to pay Charlie, but he can only do so once Alice pays him back. Alice sends a transaction to Bob after \acs{BVN}-1 verifies that Alice has sufficient funds. Bob receives the tokens and sends a transaction to Charlie through \acs{BVN}-2. However, \acs{BVN}-2 does not know about the transaction or from which \acs{BVN} it came since Alice sent her transaction through \acs{BVN}-1. Therefore, \acs{BVN}-2 will have to query all other \acs{BVN}s to verify that Bob has enough money to pay Charlie. This problem can be resolved by creating transaction flows with synthetic transactions.

After \acs{BVN}-1 has validated Alice's transaction, it sends a second transaction to \acs{BVN}-2 that says, "Deposit X tokens into Bob's account". The transaction that \acs{BVN}-1 sends to \acs{BVN}-2 is called a synthetic transaction. Now both \acs{BVN}s have a complete record of everything that has happened to their respective chains, and future transactions involving these chains will not require either \acs{BVN} to query others in the Accumulate network. To ensure that no fake synthetic transactions can be injected into the system by an external player, the \acs{DN} is used to produce cryptographic receipts for validating synthetic transactions. Thus, only valid synthetic transactions produced by the validators on a \acs{BVN} can be processed and validated.

\subsubsection{Processing Synthetic Transactions}
\begin{figure}[!h]
 \centering
 \includegraphics[scale=0.5]{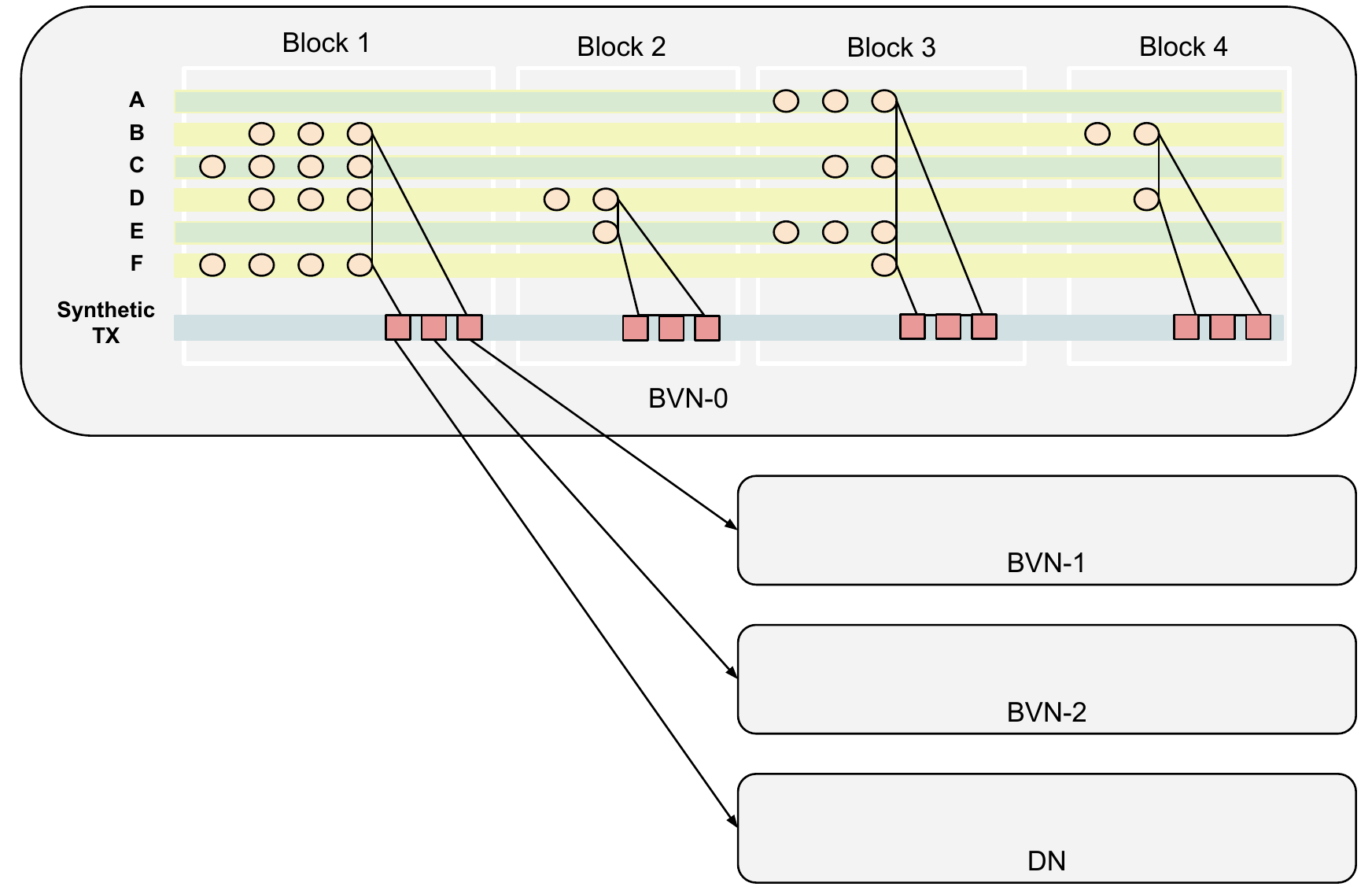}
 \caption{Synthetic transactions.}
 \label{fig:synthetic}
\end{figure}

Figure \ref{fig:synthetic} shows individual synthetic transactions between identities A-F that are being collected and sent as a bulk synthetic transaction from \acs{BVN}-0 to \acs{BVN}-1, \acs{BVN}-2, and the Directory Network (\acs{DN}). The yellow circles in rows A-F represent transactions, while those in the "Synthetic TX" row represent sets of transactions between identities A-F. The quadrangle that connects \acs{ADI}s A-F, \acs{BVN}-1, \acs{BVN}-2, and the \acs{DN} represents communication between all \acs{ADI}s hosted on \acs{BVN}-0 and all identities hosted on other \acs{BVN}s or the \acs{DN}. If the box representing \acs{BVN}-1 were opened, for example, you might find identities G-J.

When a transaction or set of transactions validates on the Accumulate network, if those transactions need to update accounts in other Root Identities, the protocol produces synthetic transactions that essentially export those updates to other \acs{BVN}s. Those updates do not require state in \acs{BVN}-0 to validate what they should do in \acs{BVN}-1, \acs{BVN}-2, or the \acs{DN}. If \acs{BVN}-0 is sending a transaction to \acs{BVN}-1 and later on to \acs{BVN}-2, \acs{BVN}-0 can operate at full speed and interact with \acs{BVN}-2 without having to wait on \acs{BVN}-1 to accept the deposit.

A real-world analogy would be the process of settling a credit card transaction. When a customer makes a purchase with their credit card, the credit card validates the transaction at the point of sale, and the goods are released by the merchant. However, the merchant does not have access to the customer's money until the transaction settles. We will imagine that the credit card platform is hosted on \acs{BVN}-0, and the merchant's bank account is hosted on \acs{BVN}-1. The customer can use their credit card for other purchases before the merchant's bank accepts their payment because the customer's credit card platform has sent a synthetic transaction to the bank broadcasting a summary of its transactions that we refer to an account balance.

\subsubsection{Summary}
For most blockchains, the transaction is the settlement. This is possible because most blockchains store the entire state of their network on full nodes. While this simplifies the process of validation, it also limits scalability. Accumulate is partitioned by design, with each identity acting as an independent blockchain and every transaction on its chain being handled by a particular \acs{BVN}. This revolutionary architecture allows the network to scale indefinitely through the addition of \acs{BVN}s, but scalability is limited by increasingly inefficient communication as the network grows. Synthetic transactions provide a means of efficient communication between \acs{BVN}s and add settlement to the blockchain.

\section{Tokenomics Model} \label{tokenomics}
Accumulate uses a burn and mint equilibrium (\acs{BME}) model for its native ACME token that trends deflationary with increasing network demand. A fixed percentage of ACME tokens are periodically minted from the unissued portion of its 500 million maximum supply and distributed to stakers and validators. Depending on transaction volume, a portion of the circulating supply will be burned to create Credits that users need to create Accumulate Digital Identifiers (\acs{ADI}s) and write data to the blockchain. Credits are non-transferable tokens with a fixed cost and no market value that disincentivize hacking, bypass regulatory requirements, and allow enterprises to budget effectively. Burned ACME tokens are returned to the unissued pool to be reissued in future blocks. Network use incentivizes staking and takes ACME out of circulation while increasing use of the network over time will decrease block rewards and put upwards pressure on the price of ACME. Thus, protocol use rather than speculation will drive the token's value. The goal of increasing the value of ACME is to incentivize users to secure the network in the absence of a traditional fee model. In this report, we will examine the dual token model of the Accumulate protocol and compare its tokenomics and value proposition to that of other blockchains.

\subsection{Deflationary Tokenomics Models}
The incentive to hold a token creates demand, while the demand of a token relative to its supply will influence its market price \cite{Freni2020}. A variety of strategies to incentivize holding and stimulate demand have been developed by blockchain protocols, the majority of which rely upon some variation of a deflationary tokenomics model. These models are briefly summarized below.

\begin{itemize}
    \item Basic deflationary model: A fixed number of tokens are created. Limited supply is expected to create demand naturally. Examples include Bitcoin (BTC), Ripple (XRP), and Solana (SOL).
    \item 	Buy back and burn: Tokens are bought back from holders and burned. This permanently removes them from the supply. Binance Coin (BNB) burns 1\% of its supply per quarter.
    \item Burn on transaction: The protocol's contract specifies a tax on transactions that burns and/or distributes the tax among its holders. Safemoon (SAFEMOON) does both.
    \item Net deflationary model: The max supply is uncapped, but the burn rate from taxes or buybacks exceeds the issuance rate. Curve (CRV) has become net deflationary at the time of writing.
\end{itemize}
 
Buy back and burn operations tend to be executed manually, which gives token issuers flexibility in the timing of buybacks and greater control over the market price. However, this often comes at the cost of decentralization. Protocols adopting this model may minimize centralization by involving a decentralized autonomous organization (\acs{DAO}) in the decision-making process.
 
Burn on transaction models are most useful for controlling the inflation of uncapped tokens. Ethereum 2.0 recently added a deflationary mechanism with its London Hard Fork that involves burning its transaction fees. However, these models are entirely dependent on transaction volume. High fees may also encourage hoarding, which could lead to undesirable price swings and excessive speculation.

Tokens with a net deflationary model generally achieve equilibrium in their later stages since buybacks from early-stage companies carry greater financial risk, and transaction rates are not high enough to offset inflation. While this model encourages activity without hoarding, equilibrium is primarily driven by speculation and may be difficult to maintain throughout market cycles.

\subsection{The Velocity Problem}
The aforementioned tokenomics models are all susceptible to the "velocity problem", which can be loosely defined as the tendency of users in a frictionless market to immediately exchange their tokens for goods and services \cite{Holden2019}. Velocity is expressed in the Equation of Exchange, which has been redefined below for cryptocurrencies and tokens:
\begin{equation}
   M\times V=P\times Q
\end{equation}
Where,\\
$M$ = Market cap of the token \\
$V$ = frequency in which a token changes hands in a given time period (i.e., velocity)\\
$P$ = Average price of goods and services purchased within this time period\\
$Q$ = number of goods and services purchased within this time period \\

The left and right sides of the equation can be interpreted as the total price of tokens spent and the total price of items bought, respectively, within a given time period. High velocity will cause an asset to be devalued, while low velocity will result in difficulty liquidating the asset. High velocity is often encountered by blockchain platforms with a "medium of exchange" token that is required to access a product or service. While the demand for this product or service may be extremely high, this does not necessarily translate to an increase in the token's value. The solution is to create a token with more economic value to holders outside of its primary utility without incentivizing them to hoard it.

\begin{table}[t] \centering
\caption{Model of ACME price given utility and speculation.}
\label{fees} \small \setlength{\tabcolsep}{4pt} \renewcommand{\arraystretch}{1.5} 
\begin{tabular}{crrrrrr} 
\hline
 &
  \multicolumn{1}{c}{} &
  \multicolumn{1}{c}{} &
  \multicolumn{4}{c}{\textbf{\footnotesize Avg. \% of Tokens burned through Fees}} \\ \cline{4-7} 
\multirow{-2}{*}{\textbf{Year}} &
  \multicolumn{1}{c}{\multirow{-2}{*}{\textbf{TPS}}} &
  \multicolumn{1}{c}{\multirow{-2}{*}{\textbf{Total Fees}}} &
  \multicolumn{1}{c}{\textbf{4.00\%}} &
  \multicolumn{1}{c}{\textbf{1.00\%}} &
  \multicolumn{1}{c}{\textbf{0.25\%}} &
  \multicolumn{1}{c}{\textbf{0.06\%}} \\ \hline
\multicolumn{1}{c|}{\textbf{1}} &
  4 &
  \$2,399,087 &
  \cellcolor[HTML]{5E913B}\$0.21 &
  \cellcolor[HTML]{5E913B}\$0.85 &
  \cellcolor[HTML]{A9D08E}\$3 &
  \cellcolor[HTML]{E2EFDA}\$14 \\
\multicolumn{1}{c|}{\textbf{2}} &
  8 &
  \$4,798,175 &
  \cellcolor[HTML]{5E913B}\$0.39 &
  \cellcolor[HTML]{A9D08E}\$1.53 &
  \cellcolor[HTML]{A9D08E}\$6 &
  \cellcolor[HTML]{E2EFDA}\$24 \\
\multicolumn{1}{c|}{\textbf{3}} &
  24 &
  \$14,394,524 &
  \cellcolor[HTML]{5E913B}\$1.06 &
  \cellcolor[HTML]{A9D08E}\$4 &
  \cellcolor[HTML]{E2EFDA}\$17 &
  \cellcolor[HTML]{A9D08E}\$67 \\
\multicolumn{1}{c|}{\textbf{4}} &
  72 &
  \$43,183,573 &
  \cellcolor[HTML]{A9D08E}\$3 &
  \cellcolor[HTML]{E2EFDA}\$12 &
  \cellcolor[HTML]{E2EFDA}\$47 &
  \cellcolor[HTML]{5E913B}\$189 \\
\multicolumn{1}{c|}{\textbf{5}} &
  288 &
  \$172,734,291 &
  \cellcolor[HTML]{A9D08E}\$11 &
  \cellcolor[HTML]{E2EFDA}\$45 &
  \cellcolor[HTML]{A9D08E}\$179 &
  \cellcolor[HTML]{5E913B}\$717 \\
\multicolumn{1}{c|}{\textbf{6}} &
  1,440 &
  \$863,671,457 &
  \cellcolor[HTML]{E2EFDA}\$54 &
  \cellcolor[HTML]{A9D08E}\$215 &
  \cellcolor[HTML]{5E913B}\$858 &
  \cellcolor[HTML]{5E913B}\$3,432 \\
\multicolumn{1}{c|}{\textbf{7}} &
  7,200 &
  \$4,318,357,287 &
  \cellcolor[HTML]{E2EFDA}\$262 &
  \cellcolor[HTML]{A9D08E}\$1,038 &
  \cellcolor[HTML]{5E913B}\$4,143 &
  \cellcolor[HTML]{5E913B}\$16,565 \\
\multicolumn{1}{c|}{\textbf{8}} &
  14,400 &
  \$8,636,714,574 &
  \cellcolor[HTML]{A9D08E}\$508 &
  \cellcolor[HTML]{5E913B}\$2,016 &
  \cellcolor[HTML]{5E913B}\$8,049 &
  \cellcolor[HTML]{5E913B}\$32,180 \\ \hline
\end{tabular}
\end{table}
 
\subsection{The Burn and Mint Equilibrium Model}
After issuing its "activation block" of 150M ACME tokens and distributing an additional 50 million ACME in a token swap for Factom's FCT holders, 300 million ACME tokens out of its 500 million maximum supply will remain in the unissued pool. Every year, 16\% of tokens in the unissued pool will be minted at intervals of approximately 1 month to compensate stakers and validators in the absence of a transaction fee.
 
ACME tokens are burned to create Credits, and Credits are used to pay for services. When ACME tokens are burned, they return to the unissued pool to be minted in future blocks. As network usage grows, the burn rate of ACME will increase, and fewer ACME tokens will remain in the circulating supply for that minting period. Since Accumulate uses a \acs{PoS} model \cite{Nguyen2019} to reward its stakers and validators, increased network use will incentivize the lockup of ACME tokens in staking pools and drive Accumulate towards a deflationary model \cite{Austin2020}.

Equilibrium is achieved due to the inverse relationship between Credits and ACME tokens, as illustrated in the following example. Assume a fixed cost of \$0.001 per Credit per kilobyte of data and a minting rate of 1 million ACME tokens. A value of \$1 per ACME token is supported with 1 billion transactions:
 
\begin{equation}
    1\ \text{mil.}\ ACME \times \frac{\$1}{ACME}\times \frac{1\  credit}{\$0.001}\times\frac{1\ tx}{1\ credit} = 1\ \text{bil.}\  tx
\end{equation}
 
If network usage increases such that 1.5 million ACME tokens are burned each month, then the circulating supply will decrease, and the value of ACME will increase. As the value of ACME increases, the number of Credits issued per ACME token will go up to support the increased demand on the network. Conversely, if network usage decreases, the circulating supply will increase and exert downwards price pressure. Thus, the price of ACME should increase linearly with the usage of the network, which directly addresses the velocity problem.

\subsection{Transition from Speculation to Utility}
All utility tokens are initially driven by speculation. As network usage grows, speculation will decrease, and the value of the token will be driven by utility. The following table models the transition from speculation to utility over time with different burn rates.

Each color represents the projected price of ACME as a function of burn rate and time, where the dark green band is least likely, and the light green band is most likely based on estimated network use and speculation. The transition from speculation to utility is represented by the leftward movement of each color over time, where utility is expected to dominate as the protocol matures. While not included in the table, a burn rate of 0\% represents zero network use, while a burn rate of 100\% represents the conversion of all circulating ACME tokens into Credits. The community would likely vote to increase the maximum supply in the latter scenario.
 
The utility is represented by the number of transactions per second (\acs{TPS}), which is expected to increase with the protocol's adoption. Note that \acs{TPS} is a function of protocol use, not the protocol capacity. The current throughput of Accumulate is 70,000 \acs{TPS} on the Testnet with a projected limit of several million \acs{TPS} over the next several years as more Block Validator Networks (\acs{BVN}s) are added to scale the network.

\subsection{Value Proposition}
The value proposition of Accumulate's \acs{BME} model is realized for enterprises who 1) want a predictable cost model, 2) cannot legally hold cryptocurrency, and 3) need the flexibility to price their own services.

\begin{itemize}
    \item Predictable costs: The price of Credits is tied to the USD, and the number of Credits required to write data to the blockchain is fixed per kilobyte of network usage. This allows enterprise users to budget their data use long-term without worrying about market conditions.
    \item Legal compliance: Some users in both the public and private sectors cannot legally hold cryptocurrency. Since Credits are non-tradeable and non-transferable, they are treated as a product rather than a security. Credits can be purchased from a third party.
    \item Flexible pricing: The Work Token Model pioneered by Augur is the only other deflationary model that addresses the velocity problem. However, it only works if service providers are offering a pure commodity. The \acs{BME} model allows service providers to set their own prices and compete with other businesses on marketing, customer service, or other variables.
\end{itemize}

\section{Use Cases}
This chapter presents example use cases regarding how Accumulate would serve humanity and bring more efficiency to everyday problems.

\subsection{Executing Multi-Party Contracts}
Thanks to its unique identity concept and key hierarchies, Accumulate is very good at executing multi-party contracts. 

\subsubsection{Managing Beneficiaries}
Some investments, securities, insurances, and assets may have beneficiaries other than the holder or primary owner of that asset, such as death beneficiaries of a health insurance. For example, the primary owner may decide where their assets are going to be distributed among several people if they pass away. The signers on the will may include family, lawyers, accountants, or business partners. While the person is alive, the signers may come and go, the owner could change lawyers. Such processes can easily be practiced with Accumulate as the owner of the asset can manage identities and accounts with lower priority levels. 

In case people lose their keys associated with a protected asset (e.g., a blockchain wallet), having implemented Accumulate may help again. With Accumulate's hierarchy of keys, the owner can backup and recover the keys held in a cold storage from a Key Page at a higher priority level. Those higher priority keys could also add or remove signatures as people enter and leave the related contract. Using a manager Key Book, which basically allows a transaction to go through only when the manager signs a transaction, the owner could assign a legal entity control of the final execution of the will.

Whenever an action is to be taken by one party (e.g., the owner or other signers), the protocol loads the Key Page mentioned in the transaction from the Key Book associated with that party. The Key Page stores the keys that are authorized to sign, particularly for the intended transaction. If the transaction is attempted to be signed by a key, not in that Key Page, the signature is rejected. Another feature of the Key Page is that it states how many keys are required at a minimum to co-sign the prompted transaction. A hypothetical Key Page may require 2 keys out of 3 total keys to sign the transaction before it becomes valid so that a quorum can be enforced for managing the assets. This relates to managing a jointly owned asset (with 3 shareholders), like a joint life insurance. If the Key Page requires all three keys to sign the contract, then a consensus will be required to take action.

Various transactions may point to different Key Pages and may require a different number of signers. In the insurance example, joint insurance owners may agree that a request for cancellation escalated by one of three owners will be enough to void the policy, but all three owners must sign to include another beneficiary (who is not an owner). In the meantime, the new beneficiary may also be requested to sign the contract to state their acknowledgment of the benefits. The beneficiary's signature would have a lower priority so that the owners can change the beneficiary at any point upon consensus. All these are easily possible with Accumulate. 

\subsection{Modeling Corporate Structures}
With its ability to manage multi-party contracts and diverse groups of digital identities, the Accumulate protocol includes all the systems required to replicate entire corporate structures. 

One of the first adoptions of blockchain technology in enterprise applications has been the use of smart contracts to automate the exchange of services and funds \cite{Ryan2017}. However, this groundbreaking innovation has been plagued by a series of large, unauthorized funds transfers that occurred due to external attacks or malfunctioning smart contracts. These incidents highlight gaps in a traditional smart contract's ability to protect an organization's assets and information while also demonstrating the risk of gate-keeping solely through a single layer of code or encryption. 
%REFERENCE NEEDED FOR FIRST SENTENCE ABOVE

Accumulate's infrastructure includes a level of redundancy and reliability that would be invaluable to enterprise users. For example, delegated networks of validators can have an active role in approving transactions. This allows companies to assign multiple validators to run concurrent algorithms verifying that the conditions of an agreement are met. Synthetic Transactions can be a valuable gatekeeping tool between two entities involved in a transaction, which prevents attackers from directly accessing any entity's funds. Transactions could also be automated via integration of URL-based addresses with external applications.

As organizations incorporate blockchain-based systems into a variety of their processes, it is important that the underlying blockchain protocols can scale as their user base grows. Not only is Accumulate able to scale its own infrastructure inexpensively, but it is also able to help companies manage changes within the dynamics of their organizations. Employee turnover (and the resulting loss of organizational knowledge), restructuring of leadership, merging or termination of departments, and acquisition or loss of customers and vendors can all be modeled and managed via Accumulate's identity and key hierarchies. The design of the Key Book allows permissions to be easily added, revoked, and upgraded while URL-based indexing ensures a clear, easily understandable organization of entities within a company.

\subsection{Token Gateways}
Wrapped tokens have been a key innovation in the blockchain space \cite{Caldarelli2022}; however, many protocols with this capability are still not able to produce and manage the data required to maintain a traded asset over time. One major challenge is managing the validation of the flow of native tokens into wrapped tokens, which is difficult because the set of responsible validators is constantly changing. With a dynamic pool of validators, complications can arise when attempting to maintain a reliable record of users' Know-Your-Customer (\acs{KYC}) and Anti-Money Laundering (\acs{AML}) details.   
%REFERENCE NEEDED ON WRAPPED TOKENS, MAYBE FIRST SENTENCE IN THE PARAGRAPH ABOVE

This is especially true when cross-chain \acs{KYC} is necessary. For example, when a wrapped token is used to buy another kind of token, the acceptor blockchain must get the \acs{KYC} information of the individual attempting the exchange in order to verify that they rightfully own the wrapped token. However, this verification step is based on the \acs{KYC} process of another protocol; if the validator that performed the original \acs{KYC} for the wrapped token has since dropped out of the protocol, there is a risk that the tokens would be invalid, with no clear proof that they were obtained legally. This kind of multi-chain authorization is beyond the capacity of most smart contracts to handle in a cost-effective manner.

Accumulate can help other protocols address this problem by documenting the gateway processes between blockchains. Accumulate's infrastructure is designed to be able to track changes over time, making it the ideal solution for preserving the audit trail from a token's creation onward. Scratch Accounts can be used to preserve important details related to \acs{KYC} at low costs off-chain while maintaining a clear, ordered record of essential data points. The reliability of the \acs{KYC} process can even be improved by assigning multiple parties to the verification process via Delegated Transactions. In addition, approval thresholds can be assigned  (ex: approval by $2/3$ of validators) to ensure that the loss of a single validator would not slow down the approval process.  

%\subsubsection{Use multi-sig to distribute custodianship of tokens, use Acc to create the set of sigs that control the multi-sig, some discussion of how to maintain custodianship over time on blockchains like BTC that do not have multi-sig. When have a change, just move tokens. It will cost only a tx.}

%\subsubsection{If a billion in gateway and costs you a single tx fee, it is negligible}

\subsection{Internet of Things}

\subsubsection{The Solutions Provided by Accumulate}
Despite the benefits of blockchain technology, its widespread application in \acs{IoT} devices is still hindered by high costs and transaction fees, limited data storage, and security concerns when data comes from a single source. Transaction fees can become prohibitively expensive when the number of \acs{IoT} devices is large or the frequency of data generation is high. Long-term storage of all the data generated by an \acs{IoT} network can still overwhelm a distributed system, limiting its growth. A single bad actor with the permission to access the data can add or delete entries after data has been collected, undermining trust in the network. 

Accumulate solves these issues by assigning an identity to each sensor, hashing and pruning the data, and allowing companies to manage their keys over time. On Accumulate, each sensor is assigned a digital identity in the form of an Accumulate Digital Identifier (\acs{ADI}), which prevents spoofing (e.g., the malicious use of duplicate sensors) and allows the user to monitor and audit data from an individual sensor in the network.  

Sensor data is hashed in a Merkle Tree, which creates efficient cryptographic proofs of data validity, while the Patricia Trie creates efficient cryptographic proofs of the current state of the system. This data structure allows an entry to be removed from the Patricia Trie while still maintaining proof that the event existed in the hashes of the Merkle Tree. From a practical point of view, there will probably be some roll-up data that an organization or utility cares about, but not for every device in the network. For example, voltage levels across every device in a power grid may be deleted after 6 months except for devices in particular locations. With Accumulate, you can prune the underlying data, extract the roll-up, delete the unnecessary data, but prove that the data the roll-up depends on existed. 

Accumulate also provides a system of robust key management. As explained in Chapter \ref{key_management}, a hierarchy of keys allows a company to assign different levels of security to the manager of an \acs{ADI} depending on the role of a key holder and the perceived value of the data. The keys to an \acs{ADI} can be managed just like signers on an account, so new \acs{ADI}s do not have to be issued with the departure or promotion of an employee or when a company shifts responsibilities over time. In addition, data on the blockchain generated by an \acs{ADI} can only be validated by someone with key access. An audit trail of their activity establishes confidence in the data across the \acs{IoT} network.

\subsubsection{ESG Scores as a Practical Application}
One \acs{IoT} use case for the Accumulate network is the generation of environmental data to provide an environmental, social, and governance (ESG) score \cite{Jang2020} for a building to quantify its environmental friendliness. Temperature, humidity, and sound level sensors may be installed in the building to continuously monitor the parameters that comprise an ESG score. Business partnerships or investments may depend on the ESG score, which is important to a growing class of sustainable investors who seek to maximize the good their investment does for society and the environment. If an investor uses \acs{IoT} data as part of their process of valuing the property, they need to make sure that the data was not fabricated by a property manager.

Accumulate provides a way to prove that the data is valid by assigning an \acs{ADI} to each sensor and maintaining an audit trail. Proving that data came from a particular sensor on a particular floor in the building may also alert the property manager to the problem that negatively impacts their ESG score. For example, a poorly insulated property on the ground floor that is generating an unexpected amount of heat. Meanwhile, pruning data after an evaluation but maintaining proof that the data existed may prove the innocence of the property manager if they are later audited. 

\subsubsection{Industrial \acs{IoT}}
Industrial \acs{IoT} (\acs{IIoT}) relates to the use of interconnected and internet-connected sensor devices throughout an industrial plant (also including facilities like farms, ports, airports, etc., in a broader sense). The sensors may continuously monitor production sites, equipment, computerized systems, air and soil parameters (e.g., temperature, humidity, wind, etc.), and the final products or yields. As a result, there may be large amounts of data produced by different departments, units, and entities related to different variables and parameters. Managing, categorizing, and classifying such data is a heavy-duty exercise for businesses. However, not all data collected by an IIoT network needs to be recorded. For example, a food manufacturer may only want to be informed of an extraordinary event that affects the safety of their product. There is also the risk of infiltration and data breaches which may cause undetectable manipulation or loss of crucial data. 

How a generic blockchain integration would contribute to this system is quite clear: with a blockchain on one end of the data flow, the data would be kept safer \cite{Cabuk2021}. However, most blockchain systems are very slow (e.g., allowing only several to a few hundred \acs{TPS}), and the unorganized data remains unorganized, making it even harder to pull when necessary. This is where Accumulate can contribute to \acs{IIoT} much better than any other blockchain system.

Accumulate allows each department (e.g., production, quality assurance), unit (e.g., shifts, teams), device (e.g., conveyors, robotic arms), area (e.g., ateliers, hangars), or any physical sector of the facility (e.g., corn farms of a hybrid plantation) to have their own unique human-readable (and internet-readable) addresses within their hierarchical accounts provided by the \acs{ADI} concept. Therefore, it is possible to categorize and classify every single data packet at the moment that it is produced within its source (e.g., sensor device). Moreover, Accumulate allows removing parts of data that do not need to be kept permanently (such as nominal temperature records) after a preset period, preventing the database from bloating. Scratch Accounts can also be used to store draft measurements that require an authorized staff to confirm (or reject).

Accumulate also allows the creation of complex and flexible authorization schemes, which is beneficial for managing and automating consensus among multiple parties. Imagine that a finished product is subject to a series of quality control tests, and it will be ready to ship only if all the tests are passed. Also imagine that some tests are done by different people or devices belonging to different departments who may change over time. Accumulate can automate consensus building through its Managed Accounts and allow the manufacturer to reassign keys without disrupting their manufacturing process. Therefore, Accumulate can increase the reliability of the production processes, provide better data security, and reduce the operational costs in many \acs{IIoT} scenarios.

\subsection{Sponsored \acs{DAO}}

\subsubsection{Origin of Decentralized Autonomous Networks}
Satoshi Nakamoto believed that repeated violations of public trust by centralized financial institutions necessitated the creation of a trust-less and decentralized peer-to-peer digital currency based on blockchain technology \cite{Nakamoto2009}. By early 2009, during a global financial crisis, many others had echoed Satoshi's sentiment as public faith in financial institutions was at its nadir. The distributed yet collaborative community of blockchain enthusiasts that began to materialize was reflected in the trust-less and decentralized nature of blockchain technology itself. As individuals, however, they lacked the power to challenge traditional financial institutions.

After smart contracts were developed by Ethereum and Proof of Stake was successfully implemented by Peercoin \cite{Zhao2021}, the organization of blockchain enthusiasts into large, distributed groups with shared financial goals became feasible. The bylaws of a decentralized organization could be written into transparent, verifiable, and publicly auditable smart contracts. Members could stake their tokens in exchange for the voting power to make critical decisions about the management of their organization, including its partnerships, technical upgrades, and treasury allocations. Even the bylaws could be changed if a consensus was reached. This collectively governed organization of stakeholders, whose operations are wholly owned and managed by its members in the absence of a central authority, is known as a decentralized autonomous organization (\acs{DAO}).
%REFERENCE NEEDED FOR PEERCOIN

\subsubsection{Growth and Legal Recognition of DAOs}
While some consider Bitcoin to be the first \acs{DAO} due to the governance mechanism of its mining network, Dash is the first modern \acs{DAO} \cite{Chistiakov2020}, launched in 2014, that provided a governance mechanism for its stakeholders and allowed them to vote on proposals that decided the future of the organization. BitShares was launched that same year as an e-commerce platform that connected merchants to customers in the absence of a central authority \cite{Bitshares2017}.
%REFERENCE NEEDED FOR BITSHARES IN PARAGRAPH ABOVE

Unfortunately, the most widely-recognized \acs{DAO} was an Ethereum-based organization simply called The \acs{DAO} that raised \$150 million from investors in 2016 before suffering from an exploit in its code that drained one-third of its treasury and ended the project a mere 5 months after launch. The funds were returned, but only at the cost of a controversial hard-fork of Ethereum \cite{Buterin2016}, which resulted in the creation of Ethereum Classic. While the attack on the \acs{DAO} sent shock waves through the blockchain community and undermined the legitimacy of DAOs for a time, development has continued, and DAOs are now considered a legal entity in the state of Wyoming, USA \cite{Ladani2021}, with more states likely to follow.
%REFERENCE NEEDED ON NEWS OF WYOMING LEGALIZING DAOS IN PARAGRAPH ABOVE

Continued innovation of DAOs and their growing recognition as legitimate financial institutions has led to their adoption by nearly every financial sector. For example, MakerDAO provides collateral-backed loans against cryptocurrency and real-world assets, JennyDAO issues fractional shares of NFTs, Nexus Mutual offers insurance, Raid Guild contributes to the gig economy, and Endaoment pools charitable contributions that are distributed by their stakeholders.

\subsubsection{Advantages and Disadvantages of DAOs}
The most compelling advantage of DAOs is their elimination of the principle-agent dilemma \cite{Beck2018}. In traditional finance, there is an inherent conflict in priorities between the principle, who invests in a fund, and the agent, who manages the fund. For example, a trader may take on excessive risk at the expense of their investors by investing in volatile assets or using highly leveraged positions with the goal of securing a large year-end bonus. Here, the agent's goal of a bonus is in potential conflict with the principal's goal of profiting from their investment. In a \acs{DAO}, however, each member contributes to the purchasing and management of an asset. Therefore, the stakeholders of a \acs{DAO} are both the agent and the principle whose priorities are aligned.
%REFERENCE NEEDED ON THE PRINCIPLE-AGENT DILEMMA IN FINANCE IN PARAGRAPH ABOVE

DAOs also benefit a stakeholder through greater efficiency in business operations, transparency in rules and operations, and autonomy in how their assets are managed. As explained below:

\begin{itemize}
    \item Efficiency: The lack of hierarchical structures inherent in traditional finance eliminates bureaucratic inefficiencies. In a \acs{DAO}, decisions are reached by community consensus after a quorum is reached.
    \item Transparency: The rules of a \acs{DAO} are embedded in smart contracts that can be publicly audited. Contract details and transactions are permanently recorded on the blockchain.
    \item Autonomy: A stakeholder has complete ownership of their investments. They can also participate in a project from inception to exit at their own discretion without facing complex regulatory hurdles.
\end{itemize}

The positive attributes of a \acs{DAO} can, unfortunately, work against the organization when things go wrong. For example, transparent code gives hackers the opportunity to simulate an attack on a virtual machine before an attack is launched on the actual network. In the case of The \acs{DAO}, a vulnerability in the code was publicly reported, and the community was called to a vote after a patch had been developed. However, the hack occurred before the voting could be completed, resulting in a loss of 11.5 million Ether \cite{Tikhomirov2018}. The decentralization of authority means that no one is held accountable for monetary losses or legal action taken against the \acs{DAO}. The involvement of non-experts in the decision-making process leaves the \acs{DAO} particularly vulnerable to lawsuits, which may be filed across multiple jurisdictions because of the decentralized nature of the \acs{DAO}. For example, a \acs{DAO} may be sued if a stakeholder contributes to a private equity fund without being an accredited investor.
%REFERENCE NEEDED ON "THE DAO" HACK IN PARAGRAPH ABOVE

\subsubsection{Sponsored \acs{DAO}}
A sponsored \acs{DAO} (\acs{SDAO}) is a theoretical organization that is collectively owned by its members but managed by an accredited institution (e.g., a bank or investment firm). The governing institution is responsible for specifying the lending criteria for borrowers, maintaining \acs{KYC}/\acs{AML} records for all stakeholders, and providing backstop liquidity for tokenized assets that are locked into Digital Special Purpose Vehicles (DSPVs) to securitize these assets and create liquidity pools for investors.

DSPVs can be multi-tranche, giving investors a choice between a lower risk investment with stable returns and a higher risk investment with variable returns. This model is being implemented by Centrifuge Inc., which recently integrated its Tinlake marketplace with MakerDAO to provide loans that are backed by real-world assets \cite{Senner2021}. Tokenized assets may also be integrated with financial oracles that aggregate real-time performance and valuation data. Sponsoring organizations with access to advanced software and financial experts who can interpret this data will help stakeholders make informed decisions about their investments.
%REFERENCE NEEDED FOR CENTRIFUGE INTEGRATING WITH MAKERDAO IN PARAGRAPH ABOVE

\subsubsection{How Accumulate Enables the Creation of an \acs{SDAO}}

Financial oracles generate a large volume of data that must be synced to the tokenized asset and credentialed by institutions with legal authority. The Accumulate protocol can process transactions with low cost, high throughput, and minimal storage requirements due to its efficient data structure. Accumulate can also integrate with traditional tech stacks, meaning that financial institutions can use the Accumulate network to manage assets without having to adapt their technology.

The Accumulate protocol's hierarchical identity and key structure enable asset management with greater flexibility and less granularity than traditional DAOs. Attestations given to stakeholders and managed by sponsoring institutions allow stakeholders to invest in different assets depending on their status (e.g., accredited investor). Complex operations like subdividing and selling a mortgaged property are possible on the Accumulate network due to its powerful identity capabilities that allow multiple key holders with different priority levels to create signature groups that can be managed over time.

To understand how this identity framework may be applied in the real world, imagine a sponsoring bank with a senior vice president (VP) in charge of agents who manage investments in the \acs{SDAO}. The bank may provide the VP with an attestation that authorizes the VP to issue their own attestations to agents who, in turn, can provide attestations to individual stakeholders. Each entity has an identity on the Accumulate network that is capable of assigning and revoking authority to those who are lower on the identity hierarchy. The bank can revoke the VP's authority just as the VP can revoke an agent's authority, and an agent can remove a stakeholder's authority if they lose accredited investor status.

Hierarchical key sets in Accumulate are also useful for managing high-value tokens. Consider an \acs{NFT} that represents a bundle of real-estate valued at \$100 million. In traditional blockchains, the loss or theft of a key can result in an irreversible loss of the \acs{NFT}. Multi-signature authorization can provide an additional layer of security with flexibility in key management. For example, adding or selling a piece of real estate might take days given time locks in the multi-signature contract and delays in acquiring the threshold number of signatures. Accumulate allows administrative keys to be maintained in cold storage while enabling routine operations with lower priority keys.

\subsubsection{Summary}
Decentralized autonomous organizations are the natural byproduct of a collective frustration with financial institutions by a decentralized collective of blockchain enthusiasts. Over time, however, DAOs began to associate with these same institutions by outsourcing certain operations, such as \acs{KYC}/\acs{AML}, to qualified organizations. With the implementation of an \acs{SDAO}, stakeholders will be able to minimize their legal exposure, make better-informed decisions, and participate in more complex financial activities while maintaining a high degree of autonomy. This type of organization is uniquely possible on the Accumulate network due to its powerful identity and key management capabilities.

\section{Conclusions}
Accumulate is a multi-chain protocol with cross-chain support that provides secure and flexible identity and key management solutions. Human-readable ADIs that adhere to W3C standards facilitate the integration of blockchain technology with servers and web-based applications. A hierarchical identity architecture with directory-like indexing makes it possible to categorize and query large datasets on-chain and replicate virtually any organizational structure. A hierarchical key architecture with prioritized Key Pages allows users to create dynamic and arbitrarily complex authorization schemes that mirror but also improve upon many financial operations. In combination, these features allow Accumulate to turn traditional applications into blockchain applications that deliver Web 3.0 technology at the level of convenience expected of Web 2.0 implementations. The engine behind this technology is Accumulate’s chain-of-chains architecture that partitions the network and provides linear scalability with unbounded TPS. Scalability is matched by the price stability provided by Accumulate’s dual-token model that allows for long-term budgeting independent of network growth.

\section{Acronyms and Glossary}
\subsection{Acronyms}
\printacronyms[heading=none]

\subsection{Glossary}
\begin{enumerate}[{}]
    \item \textbf{Accumulate:} The Accumulate network is a collection of independent chains. Each chain is managed by a hierarchy of identities known as Accumulate Digital Identifiers (\acs{ADI}s), and each identity possesses a hierarchy of keys, which allows it to participate in the execution of transactions.
   \item \textbf{ACME:} "ACME" is the symbol of a cryptographic token used in the Accumulate protocol. The ACME token, which is a traditional spendable token, like ETH and BTC and it is issued by the protocol to reward those providing services to the protocol. For example, Validators will be awarded staking fees in ACME.
   \item \textbf{Accumulate Blocks:} The Accumulate protocol defines minor blocks and major blocks. Minor blocks are once per second synchronization points of the Merkle trees. Major blocks are twice per day.
   
   \item \textbf{Accumulate Digital Identifier (\acs{ADI}):} An \acs{ADI} (also called identity) is a human-readable URL that can represent users, institutions, devices, etc. An Identity may have accounts, which hold its assets, whether those are tokens, data, or keys. An identity can be represented as: "acc://Bob". Identities provide structure on the Accumulate Blockchain and allow smart contracts, consensus building, validator networks, and enterprise-level management of digital assets.
   
   \item \textbf{ADI Data Account:} An \acs{ADI} Data Account is one of the types of accounts an \acs{ADI} can control. An \acs{ADI} data account holds data. An \acs{ADI} data account can be represented as "acc://Bob/Data". To write data, one would specify this URL.
   
   \item \textbf{ADI Token Account:} An \acs{ADI} Token Account is one of the types of accounts an \acs{ADI} can control. An \acs{ADI} token account holds tokens and can be represented as "acc://Bob/Tokens/ACME". To send or receive tokens, one would specify this URL.
   
   \item \textbf{ADI Scratch Accounts:} Scratch Accounts are identical to data accounts. However, after 2-3 weeks, its chain is compressed, and proof of the transactions is created. A proof is a logical argument used to show the truth of the operations made on the chain. This argument or proof contains less data than the Scratch Account it represents. This allows the blockchain not to be burdened with moving substantial amounts of data over the network.
   
   \item \textbf{Block Validator Network (\acs{BVN}) / Block Validator Network Node (\acs{BVNN}):} A \acs{BVN} executes transactions against records. At the end of each block, the \acs{BVN} collects Merkle DAG roots (anchors) from the Main Chain of accounts modified by one or more transactions in the block, appending them to the \acs{BVN}'s root chain. The anchor from the root chain is sent to the \acs{DN}. A \acs{BVNN} is a node with a \acs{BVN}.
   
   \item \textbf{Chain Validators/Executors:} Chain validators can be thought of as transaction executors. The actual code for executors is per-transaction (type), not per-chain. As an example, the creation of an identity would have a specified executor. The validation of the signature, for instance, is handled by the overall validator/executor.
   
   \item \textbf{Decentralized Autonomous Organization (\acs{DAO}):} A \acs{DAO} is an organization represented by rules encoded as a transparent computer program, administered by the organization members, and not influenced by a central government. A \acs{DAO}'s financial transaction record and program rules are maintained on a blockchain.
   
   \item \textbf{Credits:} Credits are a non-transferable form of payment on the Accumulate Network. Credits are created by converting ACME tokens into credits. Once the user converts ACME tokens to purchase Credits, Credits are then used for actions on the network such as the creation of an \acs{ADI}, \acs{ADI} Token Account, or updating Keys in a Key Page.
   
   \item \textbf{Directory Network (\acs{DN}):} The \acs{DN} executes transactions against certain system account, such as the ACME token issuer. The \acs{DN} also receives root chain anchors sent from \acs{BVN}s, appending them to the \acs{DN}'s \acs{BVN} anchor chain. At the end of each block, the \acs{DN} collects Merkle DAG roots (anchors) from the Main Chain of records modified by one or more transactions in the block (including the \acs{DN}'s \acs{BVN} anchor chain), appending them to the \acs{DN}'s root chain. The anchor from the root chain is sent to all of the \acs{BVN}s. 
   
   \item \textbf{Directory Network Node (DNN):} A DNN is a node within a \acs{DN}.
   
   \item \textbf{Keys:} Keys are defined as the hash of the public key for a signature.
   
   \item \textbf{Key Book:} Key Books store the Hierarchical Key Management Structure of an \acs{ADI} using Key Pages and Keys within those pages. A Key Book is an ordered set of Key Pages by priority where any Key Page can modify itself or a Page of lower priority. Modifications include adding, updating, or removing keys. An account specifies what Key Book applies to any transaction within that account.
   
   \item \textbf{Key Page:} A Key Page defines the set of Keys required to validate a transaction. A Key Page specifies one or more Keys possible and how many such Keys are required to validate a transaction. Key Pages store credits. A Key Page can be represented as so: acc://Bob/KeyPage

   \item \textbf{Layer-1 Anchoring:} An anchor is a root hash created from the latest Directory Network major block, which is then anchored into a Layer-1 blockchain such as Bitcoin.
   
   \item \textbf{Lite Data Chain:} A lite data chain is a chain that anyone can write data to, as opposed to \acs{ADI} Data Accounts which requires specified keys.
   
   \item \textbf{Lite Token Account:} A lite token account is an unstructured account that is similar to an address such as Bitcoin, a string of non-human readable characters. They are represented as "acc://<keyHash><checkSum>/<tokenUrl>". Lite Token Accounts are not associated with an identity and are not acknowledged by the Accumulate network until they have ACME tokens.
   
   \item \textbf{Major Blocks:} Every 12 hours, the transactions executed in the last 12 hours (since the previous major block) are collected into a new major block. For each record modified by a transaction in the block, a Merkle DAG root (anchor) is taken from the Transaction  Chain of the account and appended to the \acs{DN} or \acs{BVN}'s major root chain.
   
   \item \textbf{Managed Transaction:} When creating an \acs{ADI} account, you can specify an additional Key Book within the same \acs{ADI} that manages a Key Book. A manager can approve or reject a transaction submitted by an \acs{ADI}. If a transaction is rejected, a manager can suggest a new transaction on the \acs{ADI}'s Signature chain.
   
   \item \textbf{Minor Blocks:} All transactions executed in the last second are collected into a new minor block. At the end of each minor block, for each record that was modified by a transaction within the block, a Merkle DAG root (anchor) is taken from the Main Chain of the record and appended to the \acs{DN} or \acs{BVN}'s minor root chain.
   
   \item \textbf{Non-fungible Token (\acs{NFT}):} An \acs{NFT} is a non-interchangeable unit of data stored on a blockchain. Types of \acs{NFT} data units may include digital files such as photos, videos, and audio. Since each token is uniquely identifiable, NFTs differ from cryptocurrencies.
   
   \item \textbf{Signature Chains:} A Signature chain tracks transactions that have not been promoted to the Main Chain. The data that is produced in the Signature chain is pruned approximately every 2 weeks.
   
   \item \textbf{Scratch Space:} Accumulate provides scratch space on the blockchain that can be used by parties to come to a consensus but whose data availability is not retained by Accumulate forever. Scratch space allows processes to provide cryptographic proof of validation and process transactions without overburdening the blockchain.
   
   \item \textbf{sub-ADI:} An \acs{ADI} can contain another \acs{ADI}. We refer to this as a sub-ADI.
   
   \item \textbf{Synthetic Transactions:} Synthetic transactions are generated by the protocol to communicate debits and credits between \acs{BVN}s. Synthetic transactions are any transactions that are generated by the protocol. These include transactions that deposit tokens into other token accounts and refund amounts for failed transactions.
   
   \item \textbf{Token Issuance:} In Accumulate, users can create custom tokens that can be used within the Accumulate protocol.
\end{enumerate}

\balance

\section{References}

\bibliography{rsc} %You need to replace "rsc" on this line with the name of your .bib file

\providecommand*{\mcitethebibliography}{\thebibliography}
\csname @ifundefined\endcsname{endmcitethebibliography}
{\let\endmcitethebibliography\endthebibliography}{}
\begin{mcitethebibliography}{36}
\providecommand*{\natexlab}[1]{#1}
\providecommand*{\mciteSetBstSublistMode}[1]{}
\providecommand*{\mciteSetBstMaxWidthForm}[2]{}
\providecommand*{\mciteBstWouldAddEndPuncttrue}
  {\def\EndOfBibitem{\unskip.}}
\providecommand*{\mciteBstWouldAddEndPunctfalse}
  {\let\EndOfBibitem\relax}
\providecommand*{\mciteSetBstMidEndSepPunct}[3]{}
\providecommand*{\mciteSetBstSublistLabelBeginEnd}[3]{}
\providecommand*{\EndOfBibitem}{}
\mciteSetBstSublistMode{f}
\mciteSetBstMaxWidthForm{subitem}
{(\emph{\alph{mcitesubitemcount}})}
\mciteSetBstSublistLabelBeginEnd{\mcitemaxwidthsubitemform\space}
{\relax}{\relax}

\bibitem[Snow \emph{et~al.}(2018)Snow, Deery, Lu, Johnston, and
  Kirby]{Snow2018}
P.~Snow, B.~Deery, J.~Lu, D.~Johnston and P.~Kirby, \emph{Factom: Business
  Processes Secured by Immutable Audit Trails on the Blockchain}, 2018,
  \url{https://assets.website-files.com/5bca6108bae718b9ad49a5f9/5bd02670d8a1981ea62cb11f_Factom_Whitepaper_v1.2.pdf}\relax
\mciteBstWouldAddEndPuncttrue
\mciteSetBstMidEndSepPunct{\mcitedefaultmidpunct}
{\mcitedefaultendpunct}{\mcitedefaultseppunct}\relax
\EndOfBibitem
\bibitem[Merkle(1979)]{Merkle1979}
R.~C. Merkle, \emph{Method of providing digital signatures}, 1979,
  \url{https://patents.google.com/patent/US4309569}\relax
\mciteBstWouldAddEndPuncttrue
\mciteSetBstMidEndSepPunct{\mcitedefaultmidpunct}
{\mcitedefaultendpunct}{\mcitedefaultseppunct}\relax
\EndOfBibitem
\bibitem[Scottnov(2016)]{Scottnov2016}
M.~Scottnov, \emph{Gates Foundation Grant Boosts Factom's Blockchain-Based
  Medical Record Development - Bitcoin Magazine: Bitcoin News, Articles,
  Charts, and Guides}, 2016,
  \url{https://bitcoinmagazine.com/business/gates-foundation-grant-boosts-factom-s-blockchain-based-medical-record-development-1479492383}\relax
\mciteBstWouldAddEndPuncttrue
\mciteSetBstMidEndSepPunct{\mcitedefaultmidpunct}
{\mcitedefaultendpunct}{\mcitedefaultseppunct}\relax
\EndOfBibitem
\bibitem[lit(2021)]{litepaper}
\emph{ACCUMULATE Protocol Litepaper v1.0: A Universal Layer 2 Blockchain for
  DApps and DeFi}, 2021,
  \url{https://accumulatenetwork.io/Accumulate-Protocol-Litepaper-V1.0.pdf}\relax
\mciteBstWouldAddEndPuncttrue
\mciteSetBstMidEndSepPunct{\mcitedefaultmidpunct}
{\mcitedefaultendpunct}{\mcitedefaultseppunct}\relax
\EndOfBibitem
\bibitem[Monte \emph{et~al.}(2020)Monte, Pennino, and Pizzonia]{Monte2020}
G.~D. Monte, D.~Pennino and M.~Pizzonia, Proceedings of the 3rd Workshop on
  Cryptocurrencies and Blockchains for Distributed Systems, New York, NY, USA,
  2020, p. 71–76\relax
\mciteBstWouldAddEndPuncttrue
\mciteSetBstMidEndSepPunct{\mcitedefaultmidpunct}
{\mcitedefaultendpunct}{\mcitedefaultseppunct}\relax
\EndOfBibitem
\bibitem[Wood(2016)]{Wood2016}
G.~Wood, \emph{Polkadot: Vision for a Heterogeneous Multi-Chain Framework
  (Draft I)}, 2016, \url{https://polkadot.network/PolkaDotPaper.pdf}\relax
\mciteBstWouldAddEndPuncttrue
\mciteSetBstMidEndSepPunct{\mcitedefaultmidpunct}
{\mcitedefaultendpunct}{\mcitedefaultseppunct}\relax
\EndOfBibitem
\bibitem[Appel(2015)]{Appel2015}
A.~W. Appel, Verification of a Cryptographic Primitive, \emph{{ACM}
  Transactions on Programming Languages and Systems}, 2015, \textbf{37},
  1--31\relax
\mciteBstWouldAddEndPuncttrue
\mciteSetBstMidEndSepPunct{\mcitedefaultmidpunct}
{\mcitedefaultendpunct}{\mcitedefaultseppunct}\relax
\EndOfBibitem
\bibitem[Bamakan \emph{et~al.}(2020)Bamakan, Motavali, and {Babaei
  Bondarti}]{Bamakan2020}
S.~M.~H. Bamakan, A.~Motavali and A.~{Babaei Bondarti}, A survey of blockchain
  consensus algorithms performance evaluation criteria, \emph{Expert Systems
  with Applications}, 2020, \textbf{154}, 113385\relax
\mciteBstWouldAddEndPuncttrue
\mciteSetBstMidEndSepPunct{\mcitedefaultmidpunct}
{\mcitedefaultendpunct}{\mcitedefaultseppunct}\relax
\EndOfBibitem
\bibitem[Chen \emph{et~al.}(2017)Chen, Li, Luo, and Zhang]{Chen2017}
T.~Chen, X.~Li, X.~Luo and X.~Zhang, 2017 IEEE 24th International Conference on
  Software Analysis, Evolution and Reengineering (SANER), 2017, pp.
  442--446\relax
\mciteBstWouldAddEndPuncttrue
\mciteSetBstMidEndSepPunct{\mcitedefaultmidpunct}
{\mcitedefaultendpunct}{\mcitedefaultseppunct}\relax
\EndOfBibitem
\bibitem[Xiao \emph{et~al.}(2021)Xiao, Zhang, and Liu]{Xiao2021}
Y.~Xiao, P.~Zhang and Y.~Liu, Secure and Efficient Multi-Signature Schemes for
  Fabric: An Enterprise Blockchain Platform, \emph{IEEE Transactions on
  Information Forensics and Security}, 2021, \textbf{16}, 1782--1794\relax
\mciteBstWouldAddEndPuncttrue
\mciteSetBstMidEndSepPunct{\mcitedefaultmidpunct}
{\mcitedefaultendpunct}{\mcitedefaultseppunct}\relax
\EndOfBibitem
\bibitem[Dhumwad \emph{et~al.}(2017)Dhumwad, Sukhadeve, Naik, K.N., and
  Prabhu]{Dhumwad2017}
S.~Dhumwad, M.~Sukhadeve, C.~Naik, M.~K.N. and S.~Prabhu, 2017 23RD Annual
  International Conference in Advanced Computing and Communications (ADCOM),
  2017, pp. 40--43\relax
\mciteBstWouldAddEndPuncttrue
\mciteSetBstMidEndSepPunct{\mcitedefaultmidpunct}
{\mcitedefaultendpunct}{\mcitedefaultseppunct}\relax
\EndOfBibitem
\bibitem[Jung \emph{et~al.}(2002)Jung, Shishibori, Tanaka, and ichi
  Aoe]{Jung2002}
M.~Jung, M.~Shishibori, Y.~Tanaka and J.~ichi Aoe, A dynamic construction
  algorithm for the Compact Patricia trie using the hierarchical structure,
  \emph{Information Processing \& Management}, 2002, \textbf{38},
  221--236\relax
\mciteBstWouldAddEndPuncttrue
\mciteSetBstMidEndSepPunct{\mcitedefaultmidpunct}
{\mcitedefaultendpunct}{\mcitedefaultseppunct}\relax
\EndOfBibitem
\bibitem[Szpankowski(1990)]{Szpankowski1990}
W.~Szpankowski, Patricia Tries Again Revisited, \emph{J. ACM}, 1990,
  \textbf{37}, 691–711\relax
\mciteBstWouldAddEndPuncttrue
\mciteSetBstMidEndSepPunct{\mcitedefaultmidpunct}
{\mcitedefaultendpunct}{\mcitedefaultseppunct}\relax
\EndOfBibitem
\bibitem[Palm \emph{et~al.}(2018)Palm, Schelén, and Bodin]{Palm2018}
E.~Palm, O.~Schelén and U.~Bodin, 2018 Crypto Valley Conference on Blockchain
  Technology (CVCBT), 2018, pp. 31--40\relax
\mciteBstWouldAddEndPuncttrue
\mciteSetBstMidEndSepPunct{\mcitedefaultmidpunct}
{\mcitedefaultendpunct}{\mcitedefaultseppunct}\relax
\EndOfBibitem
\bibitem[Wang \emph{et~al.}(2021)Wang, Li, and Zhao]{Wang2021}
W.~Wang, X.~Li and H.~Zhao, {DCAF: Dynamic Cross-Chain Anchoring Framework
  using Smart Contracts}, \emph{The Computer Journal}, 2021\relax
\mciteBstWouldAddEndPuncttrue
\mciteSetBstMidEndSepPunct{\mcitedefaultmidpunct}
{\mcitedefaultendpunct}{\mcitedefaultseppunct}\relax
\EndOfBibitem
\bibitem[Gamage \emph{et~al.}(2020)Gamage, Weerasinghe, and Dias]{Gamage2020}
H.~T.~M. Gamage, H.~D. Weerasinghe and N.~G.~J. Dias, A Survey on Blockchain
  Technology Concepts, Applications, and Issues, \emph{{SN} Computer Science},
  2020, \textbf{1}, \relax
\mciteBstWouldAddEndPuncttrue
\mciteSetBstMidEndSepPunct{\mcitedefaultmidpunct}
{\mcitedefaultendpunct}{\mcitedefaultseppunct}\relax
\EndOfBibitem
\bibitem[Bach \emph{et~al.}(2018)Bach, Mihaljevic, and Zagar]{Bach2018}
L.~M. Bach, B.~Mihaljevic and M.~Zagar, 2018 41st International Convention on
  Information and Communication Technology, Electronics and Microelectronics
  (MIPRO), 2018, pp. 1545--1550\relax
\mciteBstWouldAddEndPuncttrue
\mciteSetBstMidEndSepPunct{\mcitedefaultmidpunct}
{\mcitedefaultendpunct}{\mcitedefaultseppunct}\relax
\EndOfBibitem
\bibitem[Zhou \emph{et~al.}(2020)Zhou, Huang, Zheng, and Bian]{Zhou2020}
Q.~Zhou, H.~Huang, Z.~Zheng and J.~Bian, Solutions to Scalability of
  Blockchain: A Survey, \emph{IEEE Access}, 2020, \textbf{8},
  16440--16455\relax
\mciteBstWouldAddEndPuncttrue
\mciteSetBstMidEndSepPunct{\mcitedefaultmidpunct}
{\mcitedefaultendpunct}{\mcitedefaultseppunct}\relax
\EndOfBibitem
\bibitem[Michelson(2021)]{Michelson2021}
K.~Michelson, \emph{Synthetic Transactions: Technical Guide}, 2021,
  \url{https://accumulatenetwork.io/2021/11/synthetic-transactions-technical-guide/}\relax
\mciteBstWouldAddEndPuncttrue
\mciteSetBstMidEndSepPunct{\mcitedefaultmidpunct}
{\mcitedefaultendpunct}{\mcitedefaultseppunct}\relax
\EndOfBibitem
\bibitem[Freni \emph{et~al.}(2020)Freni, Ferro, and Moncada]{Freni2020}
P.~Freni, E.~Ferro and R.~Moncada, 2020 IEEE Symposium on Computers and
  Communications (ISCC), 2020, pp. 1--6\relax
\mciteBstWouldAddEndPuncttrue
\mciteSetBstMidEndSepPunct{\mcitedefaultmidpunct}
{\mcitedefaultendpunct}{\mcitedefaultseppunct}\relax
\EndOfBibitem
\bibitem[Holden and Malani(2019)]{Holden2019}
R.~Holden and A.~Malani, \emph{The {ICO} Paradox: Transactions Costs, Token
  Velocity, and Token Value},  technical report, National Bureau of Economic
  Research, 2019\relax
\mciteBstWouldAddEndPuncttrue
\mciteSetBstMidEndSepPunct{\mcitedefaultmidpunct}
{\mcitedefaultendpunct}{\mcitedefaultseppunct}\relax
\EndOfBibitem
\bibitem[Nguyen \emph{et~al.}(2019)Nguyen, Hoang, Nguyen, Niyato, Nguyen, and
  Dutkiewicz]{Nguyen2019}
C.~T. Nguyen, D.~T. Hoang, D.~N. Nguyen, D.~Niyato, H.~T. Nguyen and
  E.~Dutkiewicz, Proof-of-Stake Consensus Mechanisms for Future Blockchain
  Networks: Fundamentals, Applications and Opportunities, \emph{IEEE Access},
  2019, \textbf{7}, 85727--85745\relax
\mciteBstWouldAddEndPuncttrue
\mciteSetBstMidEndSepPunct{\mcitedefaultmidpunct}
{\mcitedefaultendpunct}{\mcitedefaultseppunct}\relax
\EndOfBibitem
\bibitem[Austin \emph{et~al.}(2020)Austin, Merrill, and Rietz]{Austin2020}
T.~H. Austin, P.~Merrill and J.~Rietz, \emph{Business Information Systems
  Workshops}, Springer International Publishing, 2020, pp. 73--85\relax
\mciteBstWouldAddEndPuncttrue
\mciteSetBstMidEndSepPunct{\mcitedefaultmidpunct}
{\mcitedefaultendpunct}{\mcitedefaultseppunct}\relax
\EndOfBibitem
\bibitem[Ryan(2017)]{Ryan2017}
P.~A. Ryan, Smart Contract Relations in e-Commerce: Legal Implications of
  Exchanges Conducted on the Blockchain, \emph{Technology Innovation Management
  Review}, 2017, \textbf{7}, 10--17\relax
\mciteBstWouldAddEndPuncttrue
\mciteSetBstMidEndSepPunct{\mcitedefaultmidpunct}
{\mcitedefaultendpunct}{\mcitedefaultseppunct}\relax
\EndOfBibitem
\bibitem[Caldarelli(2022)]{Caldarelli2022}
G.~Caldarelli, Wrapping Trust for Interoperability: A Preliminary Study of
  Wrapped Tokens, \emph{Information}, 2022, \textbf{13}, \relax
\mciteBstWouldAddEndPuncttrue
\mciteSetBstMidEndSepPunct{\mcitedefaultmidpunct}
{\mcitedefaultendpunct}{\mcitedefaultseppunct}\relax
\EndOfBibitem
\bibitem[Jang \emph{et~al.}(2020)Jang, Kang, Lee, and Bae]{Jang2020}
G.-Y. Jang, H.-G. Kang, J.-Y. Lee and K.~Bae, ESG Scores and the Credit Market,
  \emph{Sustainability}, 2020, \textbf{12}, \relax
\mciteBstWouldAddEndPuncttrue
\mciteSetBstMidEndSepPunct{\mcitedefaultmidpunct}
{\mcitedefaultendpunct}{\mcitedefaultseppunct}\relax
\EndOfBibitem
\bibitem[Hakan~Altas(2021)]{Cabuk2021}
G.~D. Hakan~Altas, Umut Can~Cabuk, Data immutability and event management via
  blockchain in the Internet of things, \emph{TURK. J. OF ELECTR. ENG. COMPUT.
  SCI.}, 2021\relax
\mciteBstWouldAddEndPuncttrue
\mciteSetBstMidEndSepPunct{\mcitedefaultmidpunct}
{\mcitedefaultendpunct}{\mcitedefaultseppunct}\relax
\EndOfBibitem
\bibitem[Nakamoto(2009)]{Nakamoto2009}
S.~Nakamoto, \emph{Bitcoin: A peer-to-peer electronic cash system}, 2009,
  \url{http://www.bitcoin.org/bitcoin.pdf}\relax
\mciteBstWouldAddEndPuncttrue
\mciteSetBstMidEndSepPunct{\mcitedefaultmidpunct}
{\mcitedefaultendpunct}{\mcitedefaultseppunct}\relax
\EndOfBibitem
\bibitem[Zhao \emph{et~al.}(2021)Zhao, Yang, Luo, and Zhou]{Zhao2021}
W.~Zhao, S.~Yang, X.~Luo and J.~Zhou, 2021 The 3rd International Conference on
  Blockchain Technology, New York, NY, USA, 2021, p. 129–134\relax
\mciteBstWouldAddEndPuncttrue
\mciteSetBstMidEndSepPunct{\mcitedefaultmidpunct}
{\mcitedefaultendpunct}{\mcitedefaultseppunct}\relax
\EndOfBibitem
\bibitem[Chistiakov and Yanovich(2020)]{Chistiakov2020}
I.~Chistiakov and Y.~Yanovich, New York, NY, USA, 2020, p. 67–72\relax
\mciteBstWouldAddEndPuncttrue
\mciteSetBstMidEndSepPunct{\mcitedefaultmidpunct}
{\mcitedefaultendpunct}{\mcitedefaultseppunct}\relax
\EndOfBibitem
\bibitem[Schuh and Larimer(2017)]{Bitshares2017}
F.~Schuh and D.~Larimer, \emph{BitShares 2.0: General Overview}, 2017,
  \url{https://cryptorating.eu/whitepapers/BitShares/bitshares-general.pdf}\relax
\mciteBstWouldAddEndPuncttrue
\mciteSetBstMidEndSepPunct{\mcitedefaultmidpunct}
{\mcitedefaultendpunct}{\mcitedefaultseppunct}\relax
\EndOfBibitem
\bibitem[Buterin(2016)]{Buterin2016}
V.~Buterin, \emph{Hard Fork Completed | Ethereum Foundation Blog}, 2016,
  \url{https://blog.ethereum.org/2016/07/20/hard-fork-completed/}\relax
\mciteBstWouldAddEndPuncttrue
\mciteSetBstMidEndSepPunct{\mcitedefaultmidpunct}
{\mcitedefaultendpunct}{\mcitedefaultseppunct}\relax
\EndOfBibitem
\bibitem[Ladani(2021)]{Ladani2021}
N.~Ladani, \emph{DAOs Are Taking Over, With New Wyoming Law}, 2021,
  \url{https://finance.yahoo.com/news/daos-taking-over-wyoming-law-194516224.html}\relax
\mciteBstWouldAddEndPuncttrue
\mciteSetBstMidEndSepPunct{\mcitedefaultmidpunct}
{\mcitedefaultendpunct}{\mcitedefaultseppunct}\relax
\EndOfBibitem
\bibitem[Beck \emph{et~al.}(2018)Beck, , M\"{u}ller-Bloch, King, and
  and]{Beck2018}
R.~Beck, , C.~M\"{u}ller-Bloch, J.~L. King and and, Governance in the
  Blockchain Economy: A Framework and Research Agenda, \emph{Journal of the
  Association for Information Systems}, 2018,  1020--1034\relax
\mciteBstWouldAddEndPuncttrue
\mciteSetBstMidEndSepPunct{\mcitedefaultmidpunct}
{\mcitedefaultendpunct}{\mcitedefaultseppunct}\relax
\EndOfBibitem
\bibitem[Tikhomirov \emph{et~al.}(2018)Tikhomirov, Voskresenskaya, Ivanitskiy,
  Takhaviev, Marchenko, and Alexandrov]{Tikhomirov2018}
S.~Tikhomirov, E.~Voskresenskaya, I.~Ivanitskiy, R.~Takhaviev, E.~Marchenko and
  Y.~Alexandrov, Proceedings of the 1st International Workshop on Emerging
  Trends in Software Engineering for Blockchain, New York, NY, USA, 2018, p.
  9–16\relax
\mciteBstWouldAddEndPuncttrue
\mciteSetBstMidEndSepPunct{\mcitedefaultmidpunct}
{\mcitedefaultendpunct}{\mcitedefaultseppunct}\relax
\EndOfBibitem
\bibitem[Senner and Senner(2021)]{Senner2021}
R.~Senner and M.~Senner, \emph{Stablecoins' quest for money: who is afraid of
  credit?}, 2021, \url{https://doi.org/10.2139/ssrn.3940320}\relax
\mciteBstWouldAddEndPuncttrue
\mciteSetBstMidEndSepPunct{\mcitedefaultmidpunct}
{\mcitedefaultendpunct}{\mcitedefaultseppunct}\relax
\EndOfBibitem
\end{mcitethebibliography}
\bibliographystyle{rsc} %the RSC's .bst file

\end{document}